\documentclass[11pt,a4paper]{article}
\usepackage{jheppub}
\usepackage{stmaryrd,bbm}

\usepackage{tikz}\usetikzlibrary{matrix,arrows,shapes,decorations.markings}
\newcommand{\btp}{\begin{tikzpicture}[baseline=-5pt,scale=0.25,line width=0.7pt]}
\newcommand{\etp}{\end{tikzpicture}}

\newcommand{\rfl}{{\!re\!f\!}}
\newcommand{\orfl}{{\!op.re\!f\!}}
\newcommand{\op}{{\!op}}

\def\g{\mathfrak{g}}
\def\h{\mathfrak{h}}
\def\m{\mathfrak{m}}
\def\bb{\mathbb}
\def\wt{\widetilde}
\def\wh{\widehat}
\def\ul{\underline}
\def\dg{\dagger}
\def\xp{x^{+}}

\def\xb{x_{B}}

\newcommand\dwt[1]{\widetilde{\widetilde{#1}}}
\newcommand\dwh[1]{\widehat{\widehat{#1}}}

\title{Reflection algebra, Yangian symmetry and bound-states in AdS/CFT}

\author[a]{Niall MacKay}
\author[a,b]{and Vidas Regelskis}

\affiliation[a]{Department of Mathematics, University of York,\\Heslington, York YO10 5DD, UK}
\affiliation[b]{Institute of Theoretical Physics and Astronomy of Vilnius University,\\Go\v{s}tauto 12, Vilnius 01108, Lithuania}

\emailAdd{niall.mackay@york.ac.uk}
\emailAdd{vr509@york.ac.uk}

\abstract{We present the `Heisenberg picture' of the reflection algebra by explicitly 
constructing the boundary Yangian symmetry of an AdS/CFT superstring which ends on a 
boundary with non-trivial degrees of freedom and which preserves the full bulk Lie symmetry algebra.
We also consider the spectrum of bulk and boundary states and some automorphisms of the underlying algebras.}

\begin{document}

\maketitle

\parskip 10pt


\section{Introduction}

Recent developments in understanding the role of integrability in the 
AdS/CFT correspondence have led to new challenges requiring
novel mathematical methods and insights (see the review \cite{Review} and references
therein). The Yangian symmetry of the light-cone AdS/CFT $S$-matrix has very 
distinctive mathematical features and is still not fully understood: it appears 
in the wider context of `secret symmetries' \cite{MMT} and deformations of quantum 
affine algebras \cite{BGM}, and indeed may be pointing to new types of quantum 
group and associated quantum integrable models (see the reviews \cite{AT1,AT2} 
and references therein).

The structure of the boundary Yangian symmetry of the AdS/CFT reflection
$K$-matrices, for the various boundary conditions
that appear when superstrings end on different $D$-branes, is even more complicated 
(see \cite{DOz} for a comprehensive review of integrable boundary conditions and 
\cite{MN} for the $q$-deformed case).
The best understood boundary conditions are the $Y=0$ `giant graviton'
$D3$-brane \cite{HM} and the $Y=0$ $D7$-brane \cite{CY,MR1}, which
share the same generalized twisted Yangian structure
\cite{AN,MR2,Palla1}. Recent progress has made it clear that the D5-brane 
has a similar structure, based on achiral boundary conditions\ \cite{CRY,MR3,Reg}. 
These two cases correspond respectively to  class A and B symmetric spaces \cite{Helgason}.

In this paper we explore boundaries with non-trivial degrees of freedom 
and which preserve {\em all} of the bulk symmetry algebra,
the two best-known examples being the $Z=0$ `giant graviton' $D3$-brane
\cite{HM} and the right factor of the $Z=0$ $D7$-brane \cite{CY,MR1}.
Here there is no natural symmetric-space structure, and the problem is rather to take the
laboriously-computed reflection $K$-matrices of \cite{MR1} and find the correct mathematical context
for them. This turns out to employ deformations of the bulk Yangian charges only at 
{\em even} levels, and (what we shall call) the `Heisenberg picture' of the 
symmetry algebra, which incorporates a reflection operator into the Hopf algebra structure.
We explicitly construct this algebra and its evaluation representation.

{\bf Organization of the paper}.
In section 2 we explain the general algebraic structures presented in this paper:
we give a formal definition of our `Heisenberg picture' of
the reflection algebra, review the Yangian symmetry and its evaluation
representation, and present the level-2 twisted charges
that are crucial in building the Yangian of a boundary which preserves all of 
the bulk Lie algebra. In section 3 we specialize to AdS/CFT, reviewing the Hopf 
algebra structure of the bulk Yangian symmetry \cite{BeisY},
and then constructing the reflection Hopf algebra structure of the boundary
Yangian symmetry. In section 4 we consider the representations and
automorphisms of the bulk and boundary symmetry of the AdS/CFT superstrings.
We generalize some properties of the spectrum of states and write
down explicit forms for the bulk and boundary singlet states composed of
single magnons and two-magnon bound-states. We finish by constructing
the evaluation representation of the boundary Yangian.


\section{The underlying algebras}

We begin with a general description of the scattering
and reflection algebras and the corresponding Yangian structures in which they are embedded.

\subsection{The Schr\"{o}dinger and Heisenberg pictures}

There are two ways of describing bulk and boundary scattering matrices and the associated
symmetries, in which the emphasis is respectively on the algebra and operations within it, and
on the modules which form particle multiplets. With a slight abuse of terminology we shall call
these the `Heisenberg
picture' and the `Schr\"{o}dinger picture' respectively. In this paper we want to present the 
Heisenberg picture of the reflection algebra. We start by reviewing the scattering algebra of 
the integrable field theory and then proceed to the reflection algebra.

The physical bulk $S$-matrix is an operator acting on the tensor product
of algebra modules in the bulk
\begin{equation}
S:V_{1}(u_{1})\otimes V_{2}(u_{2})\rightarrow V_{2}(u_{2})\otimes V_{1}(u_{1}),
\end{equation}
where algebra modules $V_{i}$ are linear vector spaces and $u_{i}$ are associated continuous 
spectral parameters. Modules $V_{i}$ usually carry some discrete indices not
shown here. This is the Schr\"{o}dinger picture of the scattering,
where the states are evolving, while the representation of the symmetry
algebra is fixed. In this picture the invariance condition
under the symmetry algebra is
\begin{equation}
\left[\Delta(\bb{J}),S\right]=0,
\end{equation}
where the co-product of any Lie symmetry generator $\bb{J}$ is taken
to be trivial
\begin{equation}
\Delta(\bb{J})=\bb{J}\otimes1+1\otimes\bb{J}.
\end{equation}
Let us switch to the Heisenberg picture of the scattering, which
is described by the $S$-matrix
\begin{equation}
\mathcal{S}\in\mbox{End}\left(V_{1}(u_{1})\otimes V_{2}(u_{2})\right),
\end{equation}
equivalent to an (unbraided) $R$-matrix\footnote{The standard notation in the 
mathematical literature is to use a `check' to denote the Schr\"odinger picture, 
so that $S \cong \check{\mathcal{R}}$ and $\mathcal{S} \cong \mathcal{R}$.}.
In this picture the physical states are fixed, but the operators
are altered in the scattering process, and the invariance condition for the 
$S$-matrix reads as
\begin{equation}
\Delta^{\op}(\bb{J})\,\mathcal{S}-\mathcal{S}\,\Delta(\bb{J})=0,\label{S_inv}
\end{equation}
where $\Delta^{\op}=P\Delta P$ and $P$ is the graded permutation
operator in $V_{i}\otimes V_{j}$, i.e.
\begin{equation}
P\left(x\otimes y\right)=\left(-1\right)^{|x||y|}y\otimes x,\qquad P^{2}=I,
\end{equation}
for any $x,\, y\in V$. At the algebra level the opposite co-product
is defined as $\Delta^{\op}=\sigma\circ\Delta$, where $\sigma$ is
the two-site flip operator of the algebra.

The physical boundary scattering $K$-matrix, in the general case, is an 
intertwining operator which acts on the tensor product of bulk $V(u)$ and
boundary $V_{B}(w)$ modules
\begin{equation}
K:V(u)\otimes V_{B}(w)\rightarrow V(-u)\otimes V_{B}(w)
\end{equation}
and the condition for invariance under the symmetry algebra  is
\begin{equation}
\left[\Delta(\bb{J}),K\right]=0,\label{K_phys_inv1}
\end{equation}
where $\Delta(\bb{J})$ is a trivial co-product acting on bulk
and boundary modules. This is the Schr\"{o}dinger picture
of the reflection algebra. Here we have defined the boundary module $V_{B}(w)$ to be carrying
some boundary spectral parameter $w$. Usually, when the boundary is represented as an infinitely 
heavy state, it is only allowed (at most) to have some discrete indices, but no continuous parameter,
owing to relativistic invariance. This type of realization in the case of a 1+1-dimensional
integrable field theory with boundary was presented in \cite{Zamolodchikov1}. Here, however, 
and probably owing to the non-relativistic nature of the model, in order
to construct the boundary Yangian we must allow the boundary module to carry 
a continuous parameter.

e want to describe the Heisenberg picture of the reflection algebra.
To do so we define a $K$-matrix for reflection from
the right boundary to be a flavor-intertwining operator of the bulk
and boundary modules,
\begin{equation}
\mathcal{K}\in\mbox{End}\left(V(u)\otimes V_{B}(w)\right),
\end{equation}
with $K=\mathcal{P}\cdot\mathcal{K}$ where $\mathcal{P}$ is a parity operator 
acting on the bulk module as
\begin{equation}\label{ParOp}
\mathcal{P}:V(u)\mapsto V(-u),\quad\mathcal{P}^{2}=I,
\end{equation}
while leaving the boundary module invariant
\begin{equation}
\mathcal{P}:V_{B}(w)\mapsto V_{B}(w).\label{P_VB}
\end{equation}
Then the requirement of invariance under the reflection algebra becomes
\begin{equation}
\Delta^{\rfl}(\bb{J})\,\mathcal{K}-\mathcal{K}\,\Delta(\bb{J})=0,\label{K_inv}
\end{equation}
where we have defined the reflected co-product $\Delta^{\rfl}=\mathcal{P}\Delta\mathcal{P}$.
It is easy to move between (\ref{K_phys_inv1})
and (\ref{K_inv})
\begin{equation}
\mathcal{P}\left(\Delta^{\rfl}(\bb{J})\,\mathcal{K}-\mathcal{K}\,\Delta(\bb{J})\right)=\Delta(\bb{J})\, K-K\Delta\,(\bb{J})
\end{equation}
with the help of the parity operator. Since $\mathcal{P}$ affects the left,
bulk module only, on the Hopf algebra level the reflected co-product may be realized as
\begin{equation}
\Delta^{\rfl}(\bb{J})={\kappa}\circ\Delta(\bb{J})=\ul{\bb{J}}\otimes1+1\otimes\bb{J},\label{ref_coproduct}
\end{equation}
where we have defined
\begin{equation}
\ul{\bb{J}}:={\kappa}(\bb{J})
\end{equation}
(with the slight abuse of notation that in the co-product ${\kappa}\equiv{\kappa}\otimes1$).
Thus ${\kappa}$ is the automorphism of the symmetry algebra, 
corresponding to the effect of boundary scattering on the symmetry 
operators in the Heisenberg picture. It is analogous to the flip
operator $\sigma$ for scattering in the bulk.

Any non-trivial braiding factors will be affected by the reflection
map $\kappa$. For example a braided co-product
\begin{equation}
\Delta(\bb{J})=\bb{J}\otimes1+\mathcal{U}\otimes\bb{J},
\end{equation}
where $\mathcal{U}$ is some braiding factor, would have a reflected
partner of the form
\begin{equation}
\Delta^{\rfl}(\bb{J})=\ul{\bb{J}}\otimes1+\ul{\mathcal{U}}\otimes\bb{J}.
\end{equation}

It is easy to see that the relation between Schr\"{o}dinger and
Heisenberg pictures of the reflection algebra is given by
\begin{equation}
\kappa\circ\Delta(\bb{J})=\mathcal{P}\,\Delta(\bb{J})\,\mathcal{P},
\end{equation}
and is of the same form as for the scattering algebra,
\begin{equation}
\sigma\circ\Delta(\bb{J})=P\,\Delta(\bb{J})\, P.
\end{equation}

The reflection from a left boundary is treated very similarly. The 
flavor-intertwining $K$-matrix for reflection from the left boundary is
\begin{equation}
\mathcal{K}\in\mbox{End}\left(V_{B}(w)\otimes V(u)\right),
\end{equation}
and satisfies $\mathcal{P\,K} = P K P$ (where $\mathcal{P}$ has the same 
effect on bulk and boundary modules, but now in exchanged order). The 
requirement of invariance under the reflection algebra  becomes
\begin{equation}
\Delta^{\op}(\bb{J})\,\mathcal{K}-\mathcal{K}\,\Delta^{\orfl}(\bb{J})=0,\label{K_inv_left}
\end{equation}
where we have defined the opposite reflected co-product as
\begin{equation}
\Delta^{\orfl}(\bb{J})=\kappa\circ\Delta^{\op}(\bb{J})=\bb{J}\otimes1+1\otimes\ul{\bb{J}}.\label{op_ref_coproduct}
\end{equation}
It is related to the reflected co-product by $\Delta^{\orfl}=P\Delta^{\rfl}\, P$,
where the permutation operator acts on the tensor product of bulk
and boundary modules as
\begin{equation}
P:V(u)\otimes V_{B}(w)\rightarrow V_{B}(w)\otimes V(u).
\end{equation}
The explicit realization of the reflection automorphism $\kappa$ depends
on the corresponding field theory. We shall explicitly construct the
automorphism $\kappa$ for the AdS/CFT superstring in section
\ref{secAut}.


\subsection{Yangian symmetry and the evaluation representation}

The Yangian $\mbox{Y}(\g)$ of a Lie algebra $\g$ is a deformation of the universal
enveloping algebra of the polynomial algebra $\g[u]$. It has level-0 $\g$ generators $\bb{J}^{A}$
and level-1 $\mbox{Y}(\g)$ generators $\wh{\bb{J}}^{A}$. Their commutators have the generic form
\begin{equation}
\left[\,\bb{J}^{A},\bb{J}^{B}\right]=f_{\quad C}^{AB}\bb{J}^{C},\qquad\bigl[\,\bb{J}^{A},\wh{\bb{J}}^{B}\bigr]=f_{\quad C}^{AB}\wh{\bb{J}}^{C},
\end{equation}
 and must obey Jacobi and Serre relations
\begin{align}
\bigl[\,\bb{J}^{[A},\bigl[\,\bb{J}^{B},\bb{J}^{C]}\bigr]\bigr] & =0,\qquad\bigl[\,\bb{J}^{[A},\bigl[\,\bb{J}^{B},\wh{\bb{J}}^{C]}\bigr]\bigr]=0,\\
\bigl[\,\wh{\bb{J}}^{[A},\bigl[\,\wh{\bb{J}}^{B},\bb{J}^{C]}\bigr]\bigr] & =\frac{\alpha^3}{4}f_{\quad D}^{AG}f_{\quad E}^{BH}f_{\quad F}^{CK}f_{GHK}\bb{J}^{\{D}\bb{J}^{E}\bb{J}^{F\}}.\label{Serre}
\end{align}
The indices of structure constants $f_{\quad D}^{AB}$ are lowered
by the means of the inverse Killing-Cartan form $g_{BD}$, while $\alpha$ 
is a formal level-1 deformation parameter, and which we later set to 
unity.\footnote{The deformation parameter $\alpha$ may be regarded as 
Planck's constant when the Yangian is an auxiliary algebra rather than, 
as here, composed of quantum charges \cite{MacKay}.}

The co-product of the generators takes the form
\begin{align}
\Delta\bb{J}^{A} & =\bb{J}^{A}\otimes1+1\otimes\bb{J}^{A},\qquad\Delta\wh{\bb{J}}^{A}=
\wh{\bb{J}}^{A}\otimes1+1\otimes\wh{\bb{J}}^{A}+\frac{\alpha}{2}f_{\; BC}^{A}\bb{J}^{B}\otimes\bb{J}^{C}.
\end{align}

Finite-dimensional representations of $\mbox{Y}(\g)$ are
realized in one-parameter families, due to the `evaluation automorphism'
\begin{equation}
\tau_{v}:\mbox{Y}(\g)\rightarrow\mbox{Y}(\g)\quad\bb{J}^{A}\mapsto\bb{J}^{A}\,,\quad\wh{\bb{J}}^{A}\mapsto\wh{\bb{J}}^{A}+v\bb{J}^{A}\,,
\end{equation}
corresponding to a shift in the polynomial variable. On (the limited
set of) finite-dimensional irreducible representations of $\g$
which may be extended to representations of $\mbox{Y}(\g)$, 
these families are explicitly realized via the `evaluation map'
\begin{equation}
\mathrm{ev}_{v}:\mbox{Y}(\g)\mapsto\mbox{U}(\g)\quad\bb{J}^{A}\mapsto\bb{J}^{A}\,,\quad\wh{\bb{J}}^{A}\mapsto v\bb{J}^{A}\,,\label{ev_map}
\end{equation}
 which yields `evaluation modules', with states $\left|v\right\rangle $
carrying a spectral parameter $v$.

We shall build finite-dimensional representations of $\mbox{Y}(\g)$
by considering the tensor product of the two $\g$-modules
on which the bulk $S$-matrix acts. The action of Yangian generators
on states in the $\g$-module $V(v)$ is defined via an
`evaluation map' ansatz
\begin{equation}
\wh{\bb{J}}\left|v\right\rangle =\gamma\,(v+v_{0})\,\bb{J}\left|v\right\rangle ,\qquad\left|v\right\rangle \in V\left(v\right),\label{vJ_ansatz}
\end{equation}
 with $\gamma$ being a $\bb{C}$-number to be determined from the
field theory and $v_{0}$ being some representation parameter. The 
reflection automorphism $\kappa$ composed with the evaluation map ansatz gives
\begin{equation}
\kappa\bigl(\,\wh{\bb{J}}\,\bigr)\left|v\right\rangle =\gamma\,(-v+v_{0})\,\ul{\bb{J}}\left|v\right\rangle.
\end{equation}

The level-2 $\mbox{Y}(\g)$ charges may be defined by commuting the level-1 charges as
\begin{equation}
\dwh{\bb{J}}{}^{A}:=\frac{1}{c_{\g}}f^{A}_{\;\; BC}\,[\,\wh{\bb{J}}^{C},\wh{\bb{J}}^{B}],\label{J2_general}
\end{equation}
when the eigenvalue $c_{\g}$ of the
quadratic Casimir operator in the adjoint representation
($f_{A}^{\;\; BC}f_{CBD}=c_{\g}\,g_{AD}$) is non-vanishing. Then
\begin{equation}
[\,\wh{\bb{J}}^{A},\wh{\bb{J}}^{B}]=f^{AB}_{\quad C}\,\dwh{\bb{J}}{}^{C}+ X^{AB} ,\label{J2_comms}
\end{equation}
where the non-zero extra term $X^{AB}$ is constrained by the Serre relations
(\ref{Serre}) to satisfy $f^{[AB}_{\quad\; D} X^{C]D}=Y^{ABC}$, where $Y^{ABC}$ 
is (\ref{Serre}) (and thus a fixed cubic combination of level-0 charges) \cite{MacKay}, 
and by (\ref{J2_general}) to satisfy $f^A_{\;\;BC} X^{BC}=0$.
The evaluation map ansatz (\ref{vJ_ansatz}) for the level-2 charges is then
\begin{equation}\label{ev2_map}
\dwh{\bb{J}}\left|v\right\rangle =\gamma^{2}\,(v+v_{0})^{2}\,\bb{J}\left|v\right\rangle .
\end{equation}

\begin{figure}
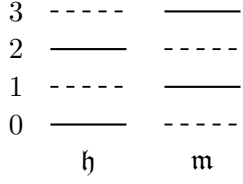

\begin{centering}
\btp
\draw[thick]  (0,0) node[left] {$0\;\;$} -- (4,0); \draw[dashed] (6,0) -- (10,0);
\draw[dashed] (0,2) node[left] {$1\;\;$} -- (4,2); \draw[thick]  (6,2) -- (10,2);
\draw[thick]  (0,4) node[left] {$2\;\;$} -- (4,4); \draw[dashed] (6,4) -- (10,4);
\draw[dashed] (0,6) node[left] {$3\;\;$} -- (4,6); \draw[thick]  (6,6) -- (10,6);
\draw (2,-2) node {$\h$} (8,-2) node {$\m$};
\etp
\par\end{centering}
\hspace{1in}%
\parbox[c]{4in}{%
\caption{The subspace structure of $\g[u]$ in case of symmetric pair decomposition 
$\g=\h\oplus\m$. Solid lines represent subspaces invariant under the involution 
$\bar\sigma$. The twisted charges of $\rm Y(\g,\h)$ live in deformations of 
the invariant subspaces.}
\label{gh}}
\end{figure}

Now let us consider a boundary module $V_{B}$ which in the general
case respects a subalgebra $\h\subset\g$ of the symmetry
algebra $\g$ of the bulk module $V$. Then integrability requires that $\h$ be 
the subalgebra invariant under an involutive automorphism $\sigma$ of $\g$, so 
that the splitting $\g=\h\oplus\m$, into subspaces of $\sigma$ eigenvalue $\pm 1$, 
forms a symmetric pair
\begin{equation}
\left[\h,\h\right]\subset\h,\qquad\left[\h,\m\right]\subset\m,\qquad\left[\m,\m\right]\subset\h.\label{symmetric_pair}
\end{equation}
Then the twisted Yangian $\mbox{Y}(\g,\h)$ is a subalgebra of $\mbox{Y}(\g)$
generated by the level-0 charges $\bb{J}^{I}$ and twisted level-1
charges
\begin{equation}
\wt{\bb{J}}^{P}:=\wh{\bb{J}}^{P}+\frac{\alpha}{4}f_{\;\; QI}^{P}\left(\bb{J}^{Q}\,\bb{J}^{I}+\bb{J}^{I}\,\bb{J}^{Q}\right),\label{twist}
\end{equation}
where $I(,J,K,...)$ run over the $\h$-indices and $P,Q(,R,...)$
over the $\m$-indices. The coproducts of the charges $\bb{J}^{I}$
and $\wt{\bb{J}}^{P}$ are
\begin{equation}
\Delta\bb{J}^{I}=\bb{J}^{I}\otimes1+1\otimes\bb{J}^{I},
\qquad\Delta\wt{\bb{J}}^{P}=\wt{\bb{J}}^{P}\otimes1+1\otimes\wt{\bb{J}}^{P}+\alpha f_{\;\; QI}^{P}\,\bb{J}^{Q}\otimes\bb{J}^{I},\label{twistcop}
\end{equation}
and satisfy the co-ideal property
\begin{equation}
\Delta\mbox{Y}(\g,\h)\in\mbox{Y}(\g)\otimes\mbox{Y}(\g,\h).\label{coideal_g_h}
\end{equation}
The charges are required to be invariant under the extension $\bar{\sigma}$ of $\sigma$
to the Yangian algebra, acting as $\bar{\sigma}(\bb{J}_{n}^{I})=(-1)^{n}\bb{J}_{n}^{I}$
and $\bar{\sigma}(\bb{J}_{n}^{P})=(-1)^{n+1}\bb{J}_{n}^{P}$,
where $n$ is the level of the charge and $\bar\sigma(\alpha)=-\alpha$, so that
$\wt{\bb{J}}^{P}$ is invariant under $\bar{\sigma}$. Thus the twisted Yangian $\mbox{Y}(\g,\h)$
is the co-ideal subalgebra of $\mbox{Y}(\g)$
 invariant under $\bar{\sigma}$ \cite{MRS}, as shown in figure \ref{gh}.

Level-2 $\mbox{Y}(\g,\h)$ charges may be generated
by the twisted level-1 charges as for (\ref{J2_general}) by
\begin{equation}
\wt{\wt{\bb{J}}}{}^{I}:=\frac{1}{c_{\g}-c_{\h}}f^{I}_{\; PQ}\,[\,\wt{\bb{J}}^{Q},\wt{\bb{J}}^{P}],\label{J2_twist}
\end{equation}
where $\wt{\bb{J}}^{P,Q}$ are defined by (\ref{twist}). It is easy to see 
that these level-2 charges satisfy the co-ideal property and are 
invariant under the involution $\bar{\sigma}$. 
We then have
\begin{equation}
[\,\wt{\bb{J}}^{P},\wt{\bb{J}}^{Q}]=f^{PQ}_{\quad I}\,	\wt{\wt{\bb{J}}}{}^{I}+ \wt{X}^{AB},
\end{equation}
however finding the deformation $\wt{\wt{\bb{J}}}_I-\wh{\wh{\bb{J}}}_I$ 
(analogous to that in (\ref{twist})) explicitly would be a rather cumbersome calculation.

Now let us turn to the case $\mathfrak{h=g}$. There is then no symmetric-pair structure, 
and no level-1 charges are conserved by the boundary. There are, however,
level-2 conserved charges, and (by extension) we write the subalgebra of charges
conserved by the boundary as $\mbox{Y}(\g,\g)$, defined to be
the $\bar{\sigma}$-invariant co-ideal subalgebra of $\mbox{Y}(\g)$
generated by the level-0 charges $\bb{J}^{A}$ and twisted
level-2 charges $\dwt{\bb{J}}{}^{A}$, where these latter are explicitly 
constructed to be invariant under the extended involution $\bar{\sigma}$ 
and to satisfy the co-ideal property
\begin{equation}
\Delta\mbox{Y}(\g,\g)\subset\mbox{Y}(\g)\otimes\mbox{Y}(\g,\g).\label{coideal_g_g}
\end{equation}
Thus, in the case when $c_{\g}\neq0$, we define $\wt{\wt{\bb{J}}}{}^{A}$
to be
\begin{eqnarray}
\dwt{\bb{J}}{}^{A} & := & \frac{1}{c_{\g}}f_{\;\; BC}^{A}\Bigl([\,\wh{\bb{J}}^{C},\wh{\bb{J}}^{B}]+\frac{\alpha}{2}f_{\;\; DE}^{C}\bb{J}^{D}[\,\wh{\bb{J}}{}^{B},\bb{J}^{E}]+\frac{\alpha}{2}f_{\;\; DE}^{B}\bb{J}^{D}[\,\bb{J}^{E},\wh{\bb{J}}{}^{C}]\Bigr)\nonumber\\
& = & \wh{\wh{\bb J}}{}^A + \frac{\alpha}{2c_{\g}}f_{\;\; BC}^{A}\Bigl(f_{\;\; DE}^{C}\bb{J}^{D}[\,\wh{\bb{J}}{}^{B},\bb{J}^{E}]+f_{\;\; DE}^{B}\bb{J}^{D}[\,\bb{J}^{E},\wh{\bb{J}}{}^{C}]\Bigr),\label{J2_gg}
\end{eqnarray}
where the last two terms ensure the co-ideal property
\begin{eqnarray}
\Delta\dwt{\bb{J}}{}^{A} & = & \dwt{\bb{J}}{}^{A}\otimes1+1\otimes\dwt{\bb{J}}{}^{A}\nonumber \\
 &  & +\frac{\alpha}{c_{\g}}f_{\;\; BC}^{A}\Bigl(f_{\;\; DE}^{C}[\,\bb{J}^{D},\wh{\bb{J}}{}^{B}]\otimes\bb{J}^{E}+f_{\;\; DE}^{B}[\,\wh{\bb{J}}{}^{C},\bb{J}^{D}]\otimes\bb{J}^{E}\Bigr)+\frac{1}
 {c_{\g}}\mathcal{O}(\alpha^{2})\\
 & \in & \mbox{Y}(\g)\otimes\mbox{Y}(\g,\g),\nonumber
\end{eqnarray}
-- i.e.\ that no level-1 charges $\wh{\bb{J}}{}^{A}$
act on the boundary -- and the order $\mathcal{O}(\alpha^{2})$ terms are
cubic in the level-0 generators and thus automatically satisfy the coideal property.

In the case when $c_{\g}=0$ (and $g_{AD}$ is degenerate), as applies in AdS/CFT, 
\eqref{J2_general} and \eqref{J2_gg} cannot be used directly 
to construct level-2 charges. One (complex) way out of this situation is to use 
Drinfeld's second realization \cite{D2}, as was done in \cite{ST}.
Alternatively one may work directly from \eqref{J2_comms}, writing it as
\begin{align}
\dwh{\bb{J}}{}^{C}\,f^{AB}_{\quad C}:=[\,\wh{\bb{J}}^{A},\wh{\bb{J}}^{B}]- X^{AB},
\end{align}
and choosing $X^{AB}$ so that this expression is non-trivial and satisfies the 
Serre relations. This approach for constructing level-2 charges for the superalgebra 
$\mathfrak{d}(1,2;\epsilon)$ (and its limiting case $\mathfrak{psu}(2|2)\ltimes\mathbb{R}^3$) 
was considered in \cite{MM2}.

However, the easiest way to construct twisted level-2 charges in this particular case 
is by observing that the term in the round brackets in the first line of \eqref{J2_gg} 
enjoys $\bar\sigma$-invariance and the co-ideal property in its own right. Thus 
(setting $\alpha$ set to unity) we define the twisted level-2 charges to be
\begin{equation}\label{J2tw}
\dwt{\bb{J}}{}^{CB} := [\,\wh{\bb{J}}^{C},\wh{\bb{J}}^{B}]+\frac{1}{2}f_{\;\; DE}^{C}\bb{J}^{D}[\,\wh{\bb{J}}{}^{B},\bb{J}^{E}]+\frac{1}{2}f_{\;\; DE}^{B}\bb{J}^{D}[\,\bb{J}^{E},\wh{\bb{J}}{}^{C}].
\end{equation}
These charges will turn out to govern the `$Z=0$' AdS/CFT boundaries, as we will show in the next section.


\section{The symmetry of the AdS/CFT superstring}

The symmetry of the excitations of the light cone string theory on
$AdS_{5}\times S^{5}$ and for the single-trace local operators in
$\mathcal{N}=4$ supersymmetric Yang-Mills gauge theory is given by
two copies of the centrally-extended Lie superalgebra \cite{BeisS}
\begin{equation}
\g=\mathfrak{psu}\left(2|2\right)\ltimes\bb{R}^{3}.\label{g}
\end{equation}


\subsection{The symmetry algebra}

Symmetry algebra $\g$ has two sets of bosonic $\mathfrak{su}(2)$ rotation generators
$\bb{R}_{a}^{\enskip b}$, $\bb{L}_{\alpha}^{\enskip\beta}$,
two sets of fermionic supersymmetry generators $\bb{Q}_{\alpha}^{\enskip a},$
$\bb{G}_{a}^{\enskip\alpha}$ and three central charges $\bb{H}$,
$\bb{C}$ and $\bb{C}^{\dg}$. The non-trivial commutation
relations are
\begin{align}
\left[\,\bb{L}_{\alpha}^{\enskip\beta},\bb{J}_{\gamma}\right]&=\delta_{\gamma}^{\beta}\bb{J}_{\alpha}-\frac{1}{2}\delta_{\alpha}^{\beta}\bb{J}_{\gamma},
&\left\{ \bb{Q}_{\alpha}^{\enskip a},\bb{Q}_{\beta}^{\enskip b}\right\}&=\epsilon^{ab}\epsilon_{\alpha\beta}\bb{C},\nonumber \\
\left[\,\bb{L}_{\alpha}^{\enskip\beta},\bb{J}^{\gamma}\right]&=-\delta_{\alpha}^{\gamma}\bb{J}^{\beta}+\frac{1}{2}\delta_{\alpha}^{\beta}\bb{J}^{\gamma},
&\left\{ \bb{G}_{a}^{\enskip\alpha},\bb{G}_{b}^{\enskip\beta}\right\}&=\epsilon^{\alpha\beta}\epsilon_{ab}\bb{C}^{\dg},\nonumber \\
\left[\,\bb{R}_{a}^{\enskip b},\bb{J}_{c}\right]&=\delta_{c}^{b}\bb{J}_{a}-\frac{1}{2}\delta_{a}^{b}\bb{J}_{c},
&\left\{ \bb{Q}_{\alpha}^{\enskip a},\bb{G}_{b}^{\enskip\beta}\right\}&=\delta_{b}^{a}\bb{L}_{\beta}^{\enskip\alpha}+\delta_{\beta}^{\alpha}\bb{R}_{b}^{\enskip a}+\delta_{b}^{a}\delta_{\beta}^{\alpha}\bb{H},\nonumber \\
\left[\,\bb{R}_{a}^{\enskip b},\bb{J}^{c}\right]&=-\delta_{a}^{c}\bb{J}^{b}+\frac{1}{2}\delta_{a}^{b}\bb{J}^{c},\label{psu(2|2)_algebra}
\end{align}
where $a,\; b,...=1,\;2$ and $\alpha,\;\beta,...=3,\;4$, and the
symbols $\bb{J}_{a}$, $\bb{J}_{\alpha}$ with lower (or upper)
indices represent any generators with the corresponding index.

The $\mathfrak{psu}(2|2)$ algebra has no matrix representation,
but the centrally extended algebra does and the representations may
traced back from the superalgebra $\mathfrak{gl}(2|2)$
using an $\mathfrak{sl}(2)$ outer-automorphism group of
the algebra.
This outer-automorphism reveals itself in the $\varepsilon\to0$ limit 
of the exceptional superalgebra $\mathfrak{d}(2,1;\varepsilon)$ 
leading to $\mathfrak{sl}(2)\ltimes\mathfrak{psu}(2|2)$
\cite{MM1}. The $\mathfrak{sl}(2)$ automorphism
transforms the supercharges of the algebra as
\begin{align}
\bb{Q}_{\;\alpha}^{\prime\enskip a}=u_{1}\bb{Q}_{\alpha}^{\enskip a}-u_{2}\varepsilon^{ab}\varepsilon_{\alpha\beta}\bb{G}_{b}^{\enskip\beta},
& \qquad\bb{G}_{\;a}^{\prime\enskip\alpha}=v_{1}\bb{G}_{a}^{\enskip\alpha}-v_{2}\varepsilon_{ab}\varepsilon^{\alpha\beta}\bb{Q}_{\beta}^{\enskip b},
\end{align}
and the central charges as
\begin{align}
\bb{C}' =u_{1}^{2}\bb{C}+u_{2}^{2}\bb{C}^{\dg}+u_{1}u_{2}\bb{H},\qquad
\bb{C}^{\prime\dg} =v_{1}^{2}\bb{C}^{\dg}+v_{2}^{2}\bb{C}+v_{1}v_{2}\bb{H},\nonumber \\
\bb{H}' =\left(u_{1}v_{1}+u_{2}v_{2}\right)\bb{H}+2u_{1}v_{1}\bb{C}+2u_{2}v_{2}\bb{C}^{\dg},\qquad\qquad
\end{align}
The parameters $u_{i}$ and $v_{i}$ satisfy the non-degeneracy requirement
$u_{1}v_{1}-u_{2}v_{2}=1$ and may be combined into a $SL(2)$ matrix
\begin{equation}\label{h_out}
h^{out}=\left(\begin{array}{cc}
u_{1} & u_{2}\\
v_{1} & v_{2}\end{array}\right).
\end{equation}
We shall be interested in the unitary representations of $\g$.
The latter requirement restricts the $SL(2)$ automorphism group to 
its real form $SU(1,1)$ upon imposing the unitarity conditions 
$v_{1}^{*}=u_{1}$ and $v_{2}^{*}=u_{2}$.
It is important to note that the outer-automorphism group leaves the
combination $\vec{\bb{H}}^{2}\equiv\bb{H}^{2}-4\bb{C}\bb{C}^{\dg}$
of the central charges invariant, i.e.\ this combination defines the
orbits of the $SL(2)$.


\subsection{Hopf algebra}

\noindent The Hopf algebra of the AdS/CFT superstring is the deformed universal
enveloping algebra $\mbox{U}(\mathfrak{psu}\left(2|2\right)\ltimes\bb{R}^{3})$
\cite{BeisY}. The deformation is defined as
\begin{equation}\label{def_Hopf}
\Delta\bb{J}^{A}=\bb{J}^{A}\otimes1+\mathcal{U}^{\left[A\right]}\otimes\bb{J}^{A},
\qquad\Delta^{\op}\,\bb{J}^{A}=\bb{J}^{A}\otimes\mathcal{U}^{\left[A\right]}+1\otimes\bb{J}^{A},
\end{equation}
where $\mathcal{U}$ is the braiding factor and its
power ${\left[A\right]}$ is defined to be
\begin{equation}
[\,\bb{R}]=[\,\bb{L}]=[\,\bb{H}]=0,\quad[\,\bb{Q}]=+1,\quad[\,\bb{G}]=-1,\quad[\,\bb{C}]=+2,\quad[\,\bb{C}^{\dg}]=-2.
\end{equation}
The co-products of the Hopf algebra are
\begin{align}
\Delta\bb{R}_{a}^{\; b}&=\bb{R}_{a}^{\; b}\otimes1+1\otimes\bb{R}_{a}^{\; b},
&\Delta\bb{Q}_{\alpha}^{\; a}&=\bb{Q}_{\alpha}^{\; a}\otimes1+\mathcal{U}^{+1}\otimes\bb{Q}_{\alpha}^{\; a},\nonumber \\
\Delta\bb{L}_{\alpha}^{\;\beta}&=\bb{L}_{\alpha}^{\;\beta}\otimes1+1\otimes\bb{L}_{\alpha}^{\;\beta},
&\Delta\bb{G}_{a}^{\;\alpha}&=\bb{G}_{a}^{\;\alpha}\otimes1+\mathcal{U}^{-1}\otimes\bb{G}_{a}^{\;\alpha},\nonumber \\
\Delta\bb{C}&=\bb{C}\otimes1+\mathcal{U}^{+2}\otimes\bb{C},
&\Delta\bb{H}&=\bb{H}\otimes1+1\otimes\bb{H}\nonumber \\
\Delta\bb{C}^{\dg}&=\bb{C}^{\dg}\otimes1+\mathcal{U}^{-2}\otimes\bb{C}^{\dg},
&\Delta\mathcal{U}&=\mathcal{U}\otimes\mathcal{U},\label{Hopf_algebra}
\end{align}
and the opposite co-products become
\begin{align}
\Delta^{\op}\,\bb{R}_{a}^{\; b}&=\bb{R}_{a}^{\; b}\otimes1+1\otimes\bb{R}_{a}^{\; b},
&\Delta^{\op}\,\bb{Q}_{\alpha}^{\; a}&=\bb{Q}_{\alpha}^{\; a}\otimes\mathcal{U}^{+1}+1\otimes\bb{Q}_{\alpha}^{\; a},\nonumber \\
\Delta^{\op}\,\bb{L}_{\alpha}^{\;\beta}&=\bb{L}_{\alpha}^{\;\beta}\otimes1+1\otimes\bb{L}_{\alpha}^{\;\beta},
&\Delta^{\op}\,\bb{G}_{a}^{\;\alpha}&=\bb{G}_{a}^{\;\alpha}\otimes\mathcal{U}^{-1}+1\otimes\bb{G}_{a}^{\;\alpha},\nonumber \\
\Delta^{\op}\,\bb{C}&=\bb{C}\otimes\mathcal{U}^{+2}+1\otimes\bb{C},
&\Delta^{\op}\,\bb{H}&=\bb{H}\otimes1+1\otimes\bb{H},\nonumber \\
\Delta^{\op}\,\bb{C}^{\dg}&=\bb{C}^{\dg}\otimes\mathcal{U}^{-2}+1\otimes\bb{C}^{\dg},
&\Delta^{\op}\,\mathcal{U}&=\mathcal{U}\otimes\mathcal{U},
\end{align}
The Hopf algebra structure becomes complete after defining the antipode
map $S$ and the co-unit map $\varepsilon$. The antipode map is the
anti-homomorphism acting on the algebra as
\begin{equation}
S\left(1\right)=1,\qquad S\left(\mathcal{U}\right)=\mathcal{U}^{-1},\qquad S(\bb{J}^{A})=-\mathcal{U}^{-\left[A\right]}\bb{J}^{A},
\end{equation}
and the co-unit map acts as
\begin{equation}
\varepsilon(1)=\varepsilon(\mathcal{U})=0,\qquad\varepsilon(\bb{J}^{A})=0.
\end{equation}
In general (off-shell), the co-products above are not co-commutative
for an arbitrary braiding factor $\mathcal{U}$, except for the the
central charge $\bb{H}$. Requiring central charges $\bb{C}$
and $\bb{C}^{\dg}$ to be co-commutative
\begin{equation}
\Delta\bb{C}=\Delta^{\op}\,\bb{C},\qquad\Delta\bb{C}^{\dg}=\Delta^{\op}\,\bb{C}^{\dg},
\end{equation}
the following relations are obtained
\begin{equation}
\bb{C}\otimes\left(1-\mathcal{U}^{+2}\right)=\left(1-\mathcal{U}^{+2}\right)\otimes\bb{C},\qquad\bb{C}^{\dg}\otimes\left(1-\mathcal{U}^{-2}\right)=\left(1-\mathcal{U}^{-2}\right)\otimes\bb{C}^{\dg},\end{equation}
leading to the constraints
\begin{equation}
\bb{C}\propto(1-\mathcal{U}^{+2}),\qquad\bb{C}^{\dg}\propto(1-\mathcal{U}^{-2}),
\end{equation}
 which fix $\mathcal{U}$ making it a non-independent generator.
One can further introduce universal proportionality coefficients $\alpha$
and $\alpha^{\dg}$
\begin{equation}
\bb{C}=\alpha(1-\mathcal{U}^{+2}),\qquad\bb{C}^{\dg}=\alpha^{\dg}(1-\mathcal{U}^{-2}),\label{C_constrains}
\end{equation}
related through the quadratic equation
\begin{equation}
\bb{C}\bb{C}^{\dg}-\alpha\bb{C}^{\dg}-\alpha^{\dg}\bb{C}=0.
\end{equation}

We want to define the reflection automorphism and the reflection Hopf
algebra using the definitions above. Let the reflection be represented
by the automorphism $\kappa$ acting on the algebra generators as
\begin{equation}
\kappa:\bb{J}^{A}\mapsto\ul{\bb{J}}^{A},\qquad\kappa:\mathcal{U}\mapsto\ul{\mathcal{U}}=\mathcal{U}^{-1},
\end{equation}
and the corresponding co-product for the reflection from the right
boundary we define to be
\begin{equation}\label{def_RHopf}
\Delta^{\rfl}\,\bb{J}^{A}=\ul{\bb{J}}^{A}\otimes1+\mathcal{U}^{-\left[A\right]}\otimes\bb{J}^{A}.
\end{equation}
Then the co-product for the reflection from the left boundary is the
reflected `op' co-product
\begin{equation}
\Delta^{\orfl}\,\bb{J}^{A}=\bb{J}^{A}\otimes\mathcal{U}^{-\left[A\right]}+1\otimes\ul{\bb{J}}^{A}.
\end{equation}

The concept of co-commutativity of the algebra charges translates
to the requirement of charge co-conservation under the reflection. 
This severely restricts the action of the automorphism $\kappa$:
\begin{equation}
\kappa(\bb{R}_{a}^{\; b})=\bb{R}_{a}^{\; b},
\qquad\kappa(\bb{L}_{\alpha}^{\;\beta})=\bb{L}_{\alpha}^{\;\beta},
\qquad\kappa(\bb{H})=\bb{H},
\end{equation}
as can easily be seen from (\ref{Hopf_algebra}). Then the only
non-trivial co-products of the reflected Hopf algebra are
\begin{align}
\Delta^{\rfl}\,\bb{Q}_{\alpha}^{\; a}&=\ul{\bb{Q}}_{\alpha}^{\; a}\otimes1+\mathcal{U}^{-1}\otimes\bb{Q}_{\alpha}^{\; a},
&\Delta^{\rfl}\,\bb{C}&=\ul{\bb{C}}\otimes1+\mathcal{U}^{-2}\otimes\bb{C},\nonumber \\
\Delta^{\rfl}\,\bb{G}_{a}^{\enskip\alpha}&=\ul{\bb{G}}_{a}^{\enskip\alpha}\otimes1+\mathcal{U}^{+1}\otimes\bb{G}_{a}^{\enskip\alpha},
&\Delta^{\rfl}\,\bb{C}^{\dg}&=\ul{\bb{C}}^{\dg}\otimes1+\mathcal{U}^{+2}\otimes\bb{C}^{\dg},
\end{align}
and in the same way the only non-trivial opposite reflected co-products are
\begin{align}
\Delta^{\orfl}\,\bb{Q}_{\alpha}^{\; a}&=\bb{Q}_{\alpha}^{\; a}\otimes\mathcal{U}^{-1}+1\otimes\ul{\bb{Q}}_{\alpha}^{\; a},
&\Delta^{\orfl}\,\bb{C}&=\bb{C}\otimes\mathcal{U}^{-2}+1\otimes\ul{\bb{C}},\nonumber \\
\Delta^{\orfl}\,\bb{G}_{a}^{\enskip\alpha}&=\bb{G}_{a}^{\enskip\alpha}\otimes\mathcal{U}^{+1}+1\otimes\ul{\bb{G}}_{a}^{\enskip\alpha},
&\Delta^{\orfl}\,\bb{C}^{\dg}&=\bb{C}^{\dg}\otimes\mathcal{U}^{+2}+1\otimes\ul{\bb{C}}^{\dg}.
\end{align}
The co-conservation requirement of the central charges
\begin{equation}\label{cons_center}
\Delta\bb{H}=\Delta^{\rfl}\,\bb{H},\qquad\Delta\bb{C}=\Delta^{\rfl}\,\bb{C},\qquad\Delta\bb{C}^{\dg}=\Delta^{\rfl}\,\bb{C}^{\dg},
\end{equation}
is clearly satisfied for $\bb{H}$, while the latter two equations
give
\begin{equation}
(\bb{C}-\ul{\bb{C}})\otimes1=(\mathcal{U}^{-2}-\mathcal{U}^{+2})\otimes\bb{C},
\qquad(\bb{C}^{\dg}-\ul{\bb{C}}^{\dg})\otimes1=(\mathcal{U}^{+2}-\mathcal{U}^{-2})\otimes\bb{C}^{\dg},
\end{equation}
which together with the constraints (\ref{C_constrains}) of the bulk charges
lead to the following constraints:
\begin{align}
\ul{\bb{C}}\otimes1&=\alpha\left(1-\mathcal{U}^{-2}\right)\otimes1,
&1\otimes\bb{C}&=1\otimes\alpha,\nonumber \\
\ul{\bb{C}}^{\dg}\otimes1&=\alpha^{\dg}\left(1-\mathcal{U}^{+2}\right)\otimes1,
&1\otimes\bb{C}^{\dg}&=1\otimes\alpha^{\dg}.\label{C^tilde-constrains}
\end{align}
The co-conservation of the opposite co-products leads to the same constraints.


\subsection{Yangian symmetry}


\paragraph{Bulk case.}

The explicit construction of the Yangian symmetry for planar AdS/CFT
was first presented in \cite{BeisY} and has been further
investigated in \cite{MM1,deLeeuw1,MMT,ST,MM2,ALT1,ALT2}. The co-products
of the Yangian charges are defined as
\begin{equation}
\Delta(\wh{\bb{J}}^{A})=\wh{\bb{J}}^{A}\otimes1+\mathcal{U}^{\left[A\right]}\otimes\wh{\bb{J}}^{A}+f_{\; BC}^{A}\,\mathcal{U}^{\left[C\right]}\bb{J}^{B}\otimes\bb{J}^{C},\label{Y_coproduct}
\end{equation}
and the opposite co-products are
\begin{equation}
\Delta^{\op}(\wh{\bb{J}}^{A})=\wh{\bb{J}}^{A}\otimes\mathcal{U}^{\left[A\right]}+1\otimes\wh{\bb{J}}^{A}+f_{\; BC}^{A}\,\bb{J}^{B}\otimes\mathcal{U}^{\left[B\right]}\,\bb{J}^{C}.\label{Y_op_coproduct}
\end{equation}
The explicit expressions of the co-products are given in appendix \ref{appA}.
It was shown in \cite{BeisY} that the co-products of the central
charges may be chosen to be co-commutative not only at the algebra
level, but also at the Yangian level. For this purpose one needs to
define the following combinations of Yangian charges
\begin{align}
\wh{\bb{H}}' =\wh{\bb{H}}+\alpha^{\dg}\bb{C}-\alpha\,\bb{C}^{\dg},\quad
\wh{\bb{C}}' =\wh{\bb{C}}+\frac{1}{2}\bb{H}(\bb{C}-2\alpha),\quad
\wh{\bb{C}}^{\dg\prime} =\wh{\bb{C}}^{\dg}-\frac{1}{2}\bb{H}(\bb{C}^{\dg}-2\alpha^{\dg}),\label{HCC_twist}
\end{align}
which we might call the `deformed central charges'. These new deformed charges
have almost-trivial co-products
\begin{align}
\Delta\wh{\bb{H}}' =\wh{\bb{H}}'\otimes1+1\otimes\wh{\bb{H}}',\quad
\Delta\wh{\bb{C}}' =\wh{\bb{C}}'\otimes1+\mathcal{U}^{+2}\otimes\wh{\bb{C}}',\quad
\Delta\wh{\bb{C}}^{\dg\prime} =\wh{\bb{C}}^{\dg\prime}\otimes1+\mathcal{U}^{-2}\otimes\wh{\bb{C}}^{\dg\prime}.
\end{align}
The co-product of the charge $\wh{\bb{H}}'$ is already co-commutative,
while the co-commutativity of $\wh{\bb{C}}'$ and $\wh{\bb{C}}^{\dg\prime}$
are ensured by imposing additional constraints
\begin{equation}
\wh{\bb{C}}'=\beta\, v_{C}\, \bb{C},\qquad \wh{\bb{C}}^{\dg\prime}=\beta\, v_{C^{\dg}}\, \bb{C}^{\dg},\label{CC_twist_ansatz}
\end{equation}
with some universal parameters $v_{C}$, $v_{C^{\dg}}$ and $\beta$.
The central charges $\wh{\bb{H}}$, $\wh{\bb{C}}$,
$\wh{\bb{C}}^{\dg}$ are also required to be co-commutative as they differ from the deformed
central charges by the central elements of the algebra only. We can also
introduce a similar ansatz
\begin{equation}
\wh{\bb{H}}'=\beta\, v_{H}\, \bb{H},\label{H_twist_ansatz}
\end{equation}
to have a complete set of expressions of the deformed central charges with $v_{H}$ 
being some universal parameter as well. 
We have not introduced or assumed any relations between the parameters
$v_{C}$, $v_{C^\dg}$ and $v_{H}$ so far, we have merely required them to be
universal\footnote{One the other hand, the universality condition is quite strong on its own.}.
We shall arrive at a set of constraints by considering the evaluation representation. 
However it is easy to see, that even at the representation level (on-shell) 
$v_{H}$ shall stay unconstrained. This is because $\bb{H}$ and $\wh{\bb{H}}'$ 
are not only co-commutative but also commutative charges.


\paragraph{Boundary case.}

We define the twisted Yangian $\rm Y(\g,\g)$ of the  $Z=0$ giant graviton to be generated
by the level-0 charges $\bb{J}^{A}$ and twisted level-2 charges
(\ref{J2tw})
\begin{align}\label{J2Z}
\wt{\wt{\bb{J}}}{}^{CB} := [\,\wh{\bb{J}}^{C},\wh{\bb{J}}^{B}\}+(-1)^{|D||B|}\frac{1}{2}f_{\;\; DE}^{C}\bb{J}^{D}[\,\wh{\bb{J}}{}^{B},\bb{J}^{E}\}+(-1)^{|D||C|}\frac{1}{2}f_{\;\; DE}^{B}\bb{J}^{D}[\,\bb{J}^{E},\wh{\bb{J}}{}^{C}\},
\end{align}
where $[\,,\}$ represents a graded commutator and $(-1)^{|D||B|}$ 
with $(-1)^{|D||C|}$ are grade factors. We do not present the explicit 
form of these level-2 charges and their co-products, ---they are complex 
but not very illuminating, and are easily obtained with the help of 
level-1 charges and their co-products, as detailed in appendix \ref{appA}, 
while for finding the expressions of the reflected co-products one has 
to use \eqref{def_RHopf} together with
\begin{align}
\Delta^\rfl(\wh{\bb{J}}^{A})=\ul{\wh{\bb{J}}}^{A}\otimes1+\mathcal{U}^{-[A]}\otimes\wh{\bb{J}}^{A}+f_{\; BC}^{A}\,\mathcal{U}^{-[C]}\ul{\bb{J}}^{B}\otimes\bb{J}^{C}.\label{Y_refcoproduct}
\end{align}

In general case, the twisted charges \eqref{J2Z} are linear combinations 
of level-2 and level-0 charges, and one loses track of the central 
elements of the algebra. For example, the charges defined by
\begin{equation}\label{C2nc}
\dwh{\bb{C}}\,' = \epsilon^{\alpha\beta}\epsilon_{ab} \{\wh{\bb{Q}}_\alpha^{\;\;a}, \wh{\bb{Q}}_\beta^{\;\;b} \},\qquad
\dwh{\bb{C}}\,^{\dg\prime} = \epsilon^{ab}\epsilon_{\alpha\beta} \{\wh{\bb{G}}_a^{\;\;\alpha}, \wh{\bb{G}}_b^{\;\;\beta} \}
\end{equation}
in contrast to $\wh{\bb{C}}$, $\wh{\bb{C}}^\dg$ and $\bb{C}$, $\bb{C}^\dg$ 
re not central, but rather are shifted from the center by a level-0 
deformation, and so are not co-commutative. Thus the twisted charges 
$\dwt{\bb{C}}\,'$ and $\dwt{\bb{C}}\,^{\dg\prime}$ that one would 
obtain using \eqref{J2Z} would not be co-conserved. However, this 
shift may be easily undone in Drinfeld's second realization \cite{D2}. 
It was shown in \cite{ST} that the charges\footnote{Our notation 
agrees with that of \cite{ALT2}.}
\begin{equation}
\dwh{\bb{C}} = \{i\wh{\bb{Q}}_4^{\;1}-w_2, i\wh{\bb{Q}}_3^{\;2}-w_3 \},\qquad
\dwh{\bb{C}}\,^\dg = \{i\wh{\bb{G}}_1^{\;4}-z_2, i\wh{\bb{G}}_2^{\;3}-z_3 \},
\end{equation}
are central. Here
\begin{align}
w_2 &= - \frac{1}{4} \{i\bb{Q}_4^{\;1},\kappa_{2,0}\} + \frac{3 i}{4} \bb{Q}^{\;1}_3 \bb{L}^{\;3}_4 -\frac{i}{4} \bb{R}^{\;1}_2 \bb{Q}^{\;2}_4 - \frac{i}{4} \bb{Q}^{\;2}_4 \bb{R}^{\;1}_2 - \frac{i}{4} \bb{L}^{\;3}_4 \bb{Q}^{\;1}_3 - \frac{i}{2} \bb{G}^{\;3}_2 \bb{C},\nonumber\\
w_3 &= - \frac{1}{4} \{i\bb{Q}_3^{\;2},\kappa_{3,0}\} - \frac{i}{4} \bb{Q}^{\;1}_3 \bb{R}^{\;2}_1 +\frac{3 i}{4} \bb{R}^{\;2}_1 \bb{Q}^{\;1}_3 - \frac{i}{4} \bb{Q}^{\;2}_4 \bb{L}^{\;4}_3 - \frac{i}{4} \bb{L}^{\;4}_3 \bb{Q}^{\;2}_4 - \frac{i}{2} \bb{G}^{\;4}_1 \bb{C},\nonumber\\
z_2 &= - \frac{1}{4} \{i\bb{G}_1^{\;4},\kappa_{2,0}\} - \frac{i}{4} \bb{G}^{\;3}_1 \bb{L}^{\;4}_3 +\frac{3 i}{4} \bb{L}^{\;4}_3 \bb{G}^{\;3}_1 - \frac{i}{4} \bb{G}^{\;4}_2 \bb{R}^{\;2}_1 - \frac{i}{4} \bb{R}^{\;2}_1 \bb{G}^{\;4}_2 - \frac{i}{2} \bb{Q}^{\;2}_3 \bb{C}^\dag,\nonumber\\
z_3 &= - \frac{1}{4} \{i\bb{G}_2^{\;3},\kappa_{3,0}\} - \frac{i}{4} \bb{G}^{\;4}_2 \bb{L}^{\;3}_4 - \frac{i}{4} \bb{L}^{\;3}_4 \bb{G}^{\;4}_2 + \frac{3 i}{4} \bb{G}^{\;3}_1 \bb{R}^{\;1}_2 - \frac{i}{4} \bb{R}^{\;1}_2 \bb{G}^{\;3}_1 - \frac{i}{2} \bb{Q}^{\;1}_4 \bb{C}^\dag,
\end{align}
and
\begin{align}
\kappa_{2,0} = -\bb{R}_1^{\;1} + \bb{L}^{\;3}_3 - \frac{1}{2}\bb{H},\qquad
\kappa_{3,0} = \bb{R}_1^{\;1} - \bb{L}^{\;3}_3 - \frac{1}{2}\bb{H}.
\end{align}
Then it is easy to see that the twisted charges
\begin{equation}
\dwt{\bb{C}} = \dwt{\bb{C}}\,'+\{w_2,w_3\},\qquad
\dwt{\bb{C}}\,^\dg = \dwt{\bb{C}}\,'^\dg+\{z_2,z_3\},
\end{equation}
are the central elements of $\rm Y(\g,\g)$ and must be co-conserved, 
i.e.\ the following constraints must hold:
\begin{align}\label{C2cons}
\Delta\dwt{\bb{C}}=\Delta^\rfl\,\dwt{\bb{C}},\qquad
\Delta\dwt{\bb{C}}\,^\dg=\Delta^\rfl\,\dwt{\bb{C}}\,^\dg.
\end{align}
Due to their complicated structure it is hard to show that these 
constraints hold at the algebraic level. 
For our purposes it will be enough to consider these constraints on-shell, 
as this is sufficient to write the evaluation map ansatz for the level-2 
charges acting on the boundary module and -- crucially -- to 
determine the associated spectral parameter related to the boundary states.


\section{Representations, states and dynamics of the AdS/CFT superstring}

\noindent The centrally-extended $\mathfrak{psu}(2|2)$
algebra has several different types of finite-dimensional representations.
The most relevant representations for AdS/CFT superstrings are
called long (typical) and short (atypical). There are also anomalous
(singlet and adjoint) representations. See \cite{BeisS} and \cite{BeisB}
for a comprehensive review and details. A tensor product of two
short representations generically yields a sum of long multiplets.
The long representations are generically irreducible, but become reducible
for some special eigenvalues of the central charges. We shall briefly
review the decomposition of the tensor product of two fundamental
representations and the tensor product of two 2-particle bound-state
representations, as this will be important to us later on.

The fundamental excitations (asymptotic states) of the superstring
transform in the 4-d(imensional) short (fundamental) representation $\boxslash$.
The tensor product of two fundamental representations
gives a 16-d irreducible long multiplet. This is the smallest
long representation. At the special points (corresponding to special
eigenvalues of the central charges) one may decompose the 16-d
long multiplet into two 8-d short representations (totally symmetric
and totally antisymmetric), or into two singlets (corresponding to the fundamental
singlet state of the spectrum) and a minimal 14-d
adjoint, which may further be reduced to $\left(3+2\times4+3\right)$-d
totally symmetric multiplets. We shall mainly focus on generic totally symmetric
short and singlet representations, where the interesting physical
states (magnons and their bound-states) of the AdS/CFT superstring
live.

Two-particle bound-states live in an 8-d totally symmetric short representation 
$\boxslash\!\boxslash$. A tensor product of two such representations decomposes into
a sum of two long, 16-d and 48-d, representations. This tensor product is very important
in the scattering theory we shall be considering, as it is the simplest representation
for which the Lie algebra is not enough to determine all the scattering coefficients and 
additional contraints are required \cite{AF1}.
Thus this representation serves as the most simple
non-trivial test of Yangian symmetry.

A general $l$-magnon bound state is described by a short totally symmetric
representation  $\boxslash\negmedspace\negmedspace\boxslash\!...\boxslash$.
The dimension of the representation is $2l|2l$ and it may be neatly
realized as degree-$l$ monomials on a graded vector space with
the basis $\omega_{1}$, $\omega_{2}$, $\theta_{3}$, $\theta_{4}$,
where $\omega_{a}$ and $\theta_{\alpha}$ are bosonic and fermionic
variables respectively \cite{AF1}.

In this representation the centrally-extended $\mathfrak{psu}(2|2)$
generators are realized as the differential operators
\begin{align}
\bb{R}_{a}^{\enskip b}&=\omega_{a}\frac{\partial}{\partial\omega_{b}}-\frac{1}{2}\delta_{a}^{b}\,\omega_{c}\frac{\partial}{\partial\omega_{c}},
&\bb{L}_{\alpha}^{\enskip\beta}&=\theta_{\alpha}\frac{\partial}{\partial\theta_{\beta}}-\frac{1}{2}\delta_{\alpha}^{\beta}\,\theta_{\gamma}\frac{\partial}{\partial\theta_{\gamma}},\nonumber \\
\bb{Q}_{\alpha}^{\enskip a}&=a\,\theta_{\alpha}\frac{\partial}{\partial\omega_{a}}+b\,\epsilon^{ab}\epsilon_{\alpha\beta}\omega_{b}\frac{\partial}{\partial\theta_{\beta}},
&\bb{G}_{a}^{\enskip\alpha}&=c\,\epsilon_{ab}\epsilon^{\alpha\beta}\theta_{\beta}\frac{\partial}{\partial\omega_{b}}+d\,\omega_{a}\frac{\partial}{\partial\theta_{\alpha}},\nonumber \\
\bb{C}&=ab\left(\omega_{a}\frac{\partial}{\partial\omega_{a}}+\theta_{\alpha}\frac{\partial}{\partial\theta_{\alpha}}\right),
&\bb{C}^{\dg}&=cd\left(\omega_{a}\frac{\partial}{\partial\omega_{a}}+\theta_{\alpha}\frac{\partial}{\partial\theta_{\alpha}}\right),\nonumber \\
\bb{H}&=\left(ad+bc\right)\left(\omega_{a}\frac{\partial}{\partial\omega_{a}}+\theta_{\alpha}\frac{\partial}{\partial\theta_{\alpha}}\right).
\end{align}
The corresponding vector space is denoted as $\mathcal{V}^{l}(p,\zeta)$,
where $p$ and $\zeta$ are complex parameters of the representation
and correspond to the momentum and the phase of an individual magnon
in the spin chain. The parameters $a,\: b,\; c,\; d$ are convenient
representation labels of the states, which we shall discuss in the
next paragraph. It is worth noting that the $SL(2)$ automorphism 
\eqref{h_out} may be realized as a shift
\begin{equation}
h^{out}:\left(\begin{array}{cc}
a & b\\
c & d\end{array}\right)\mapsto\left(\begin{array}{cc}
u_{1} & u_{2}\\
v_{1} & v_{2}\end{array}\right)\left(\begin{array}{cc}
a & b\\
c & d\end{array}\right)
\end{equation}
of the representation labels.


\subsection{Superstrings in the bulk}

Let us review the general concepts of the AdS/CFT superstring in the
bulk: the representation, $S$-matrix and  spectrum of the states.

\paragraph{The bulk representation.}

A convenient parametrization of the representation labels of the 
$l$-particle bound-states in the bulk is \cite{BeisS,AF1}
\begin{equation}
a=\sqrt{\frac{g}{2l}}\eta,\quad b=\sqrt{\frac{g}{2l}}\frac{i\zeta}{\eta}\left(\frac{x^{+}}{x^{-}}-1\right),\quad 
 c=-\sqrt{\frac{g}{2l}}\frac{\eta}{\zeta x^{+}},\quad d=-\sqrt{\frac{g}{2l}}\frac{x^{+}}{i\eta}\left(\frac{x^{-}}{x^{+}}-1\right),\label{abcd}
\end{equation}
where $g$ is a coupling constant, $\zeta={\rm e}^{2i\xi}$ is the
magnon phase and $x^{\pm}$ are the spectral parameters ($e^{ip}=\frac{x^{+}}{x^{-}}$)
respecting the mass-shell (multiplet-shortening) condition of the
$l$-magnon bound state,
\begin{equation}
x^{+}+\frac{1}{x^{+}}-x^{-}-\frac{1}{x^{-}}=i\frac{2l}{g}.\label{shortening}
\end{equation}
Unitarity requires $\eta={\rm e}^{i\xi}{\rm e}^{i\frac{\varphi}{2}}\sqrt{i\left(x^{-}-x^{+}\right)}$,
where the arbitrary phase factor ${\rm e}^{i\varphi}$ reflects the
freedom in choosing $x^{\pm}$. We shall set the phase to be $\varphi=p/2$.
This parametrization is in the so-called non-local (string) basis, where the
non-trivial braiding factor $\cal{U}$ of the corresponding Hopf algebra
is absorbed into the parametrization of labels, thus introducing the phase
$\zeta$. By setting $\zeta=1$ one recovers the local (spin-chain)
basis of the $\cal{U}$-deformed Hopf algebra \eqref{def_Hopf}.

The rapidity of the magnon in the $x^\pm$ parametrization is defined to be
\begin{equation}
u=x^{+}+\frac{1}{x^{+}}-i\frac{l}{g},\label{rapidity}
\end{equation}
and the eigenvalues of the central charges of the $l$-magnon bound state
are expressed as
\begin{eqnarray}
 & C_{l}=l\, ab=\frac{i}{2}g\left({\rm e}^{ip}-1\right){\rm e}^{2i\xi},\qquad C_{l}^{\dg}=l\, 
 cd=-\frac{i}{2}g\left({\rm e}^{-ip}-1\right){\rm e}^{-2i\xi},\nonumber \\
 & H_{l}=l\left(ad+bc\right)=\sqrt{l^{2}+4g^{2}\sin^{2}\frac{p}{2}},
\end{eqnarray}
where the braiding factor is set to $\mathcal{U}^{2}=\frac{x^{+}}{x^{-}}$.

\paragraph{The S-matrix.}

\begin{figure}[h]\begin{center}\epsfig{file=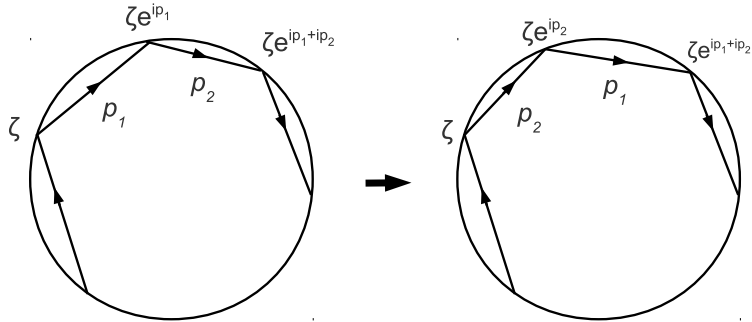,width=10cm}\caption[Scattering]
{Scattering of two well-separated magnons with momenta $p_1$ and
$p_2$ living on a long string and $\zeta$ being the reference point.}\label{fig_smatrix}
\end{center}\end{figure}

The $S$-matrix in superspace is
realized as an intertwining differential operator acting on the tensor
product of two algebra modules and may be represented as (see figure
\ref{fig_smatrix})
\begin{equation}
\mathcal{S}(p_{1},p_{2})=\sum_{i}a_{i}(p_{1},p_{2})\,\Lambda_{i},
\end{equation}
where $\Lambda_{i}$ span a complete basis of differential operators
invariant under the $\mathfrak{su}(2)\oplus\mathfrak{su}(2)$
algebra and $a_{i}(p_{1},p_{2})$ are $S$-matrix coefficients.
The exact expressions of $\Lambda_{i}$ for various $S$-matrices
are given in \cite{AF1}. The coefficients of the fundamental $S$-matrix
may be found by demanding its invariance (\ref{S_inv}) under the
symmetry algebra $\g$ (\ref{g}). However, the symmetry
algebra alone is not enough to fix uniquely \textit{all} coefficients
of the bound-state $S$-matrix and one additionally needs to use either the
Yang-Baxter equation \cite{AF1,MR1} or Yangian symmetry
\cite{BeisY,deLeeuw1,ALT1}.

\paragraph{The spectrum and the special points.} 

The spectrum of excitations of the light cone superstrings in AdS/CFT
includes fundamental and bound-state excitations and the singlet (mirror)states \cite{BeisB,DO}.

\begin{figure}\begin{center}
{\epsfig{file=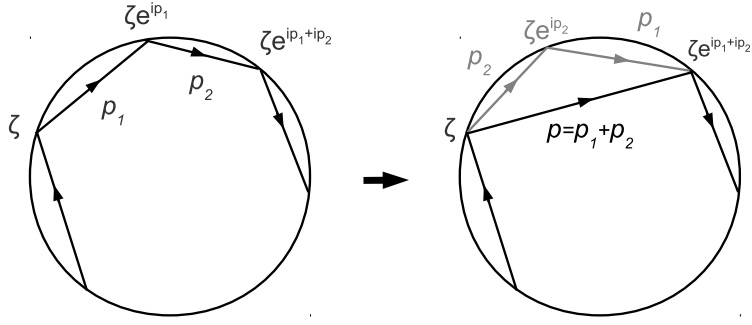,width=10cm}\caption[Bound-state]
{Construction of a two-particle bound-state as a pole $x_{1}^{-}=x_{2}^{+}$
in the $S$-matrix. The bound-state which emerges from the $S$-matrix has momentum
$e^{i p} = e^{i p_1 + i p_2} = \frac{x_{1}^{+}}{x_{1}^{-}}
\frac{x_{2}^{+}}{x_{2}^{-}} = \frac{x_{2}^{+}}{x_{1}^{-}} \equiv \frac{y^{+}}{y^{-}}$.}\label{fig_bs}}
\end{center}\end{figure}

Bound states appear as poles $x_{2}^{-}=x_{1}^{+}$ in the $S$-matrix,
projecting it to the totally symmetric part of the tensor product $V(p_{1})\otimes V(p_{2})$
and signalling the presence of the tower of multi-particle bound-states.
The new bound state emerging from the $S$-matrix has the spectral
parameters set to $y^{+}=x_{2}^{+}$ and $y^{-}=x_{1}^{-}$ (see figure \ref{fig_bs}).

\begin{figure}\begin{center}
{\epsfig{file=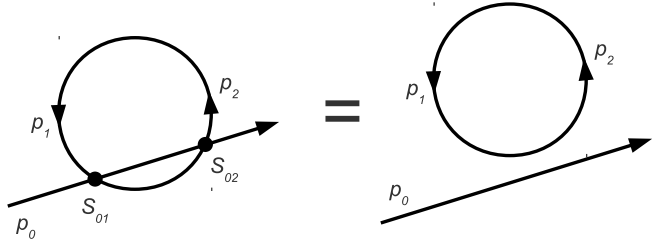,width=8.5cm}\caption[Scattering]
{Scattering of a magnon with a singlet state. This scattering is
considered to be trivial and is required to be equivalent to no
scattering at all.}
\label{fig_buble}}
\end{center}\end{figure}

\begin{figure}\begin{center}
{\epsfig{file=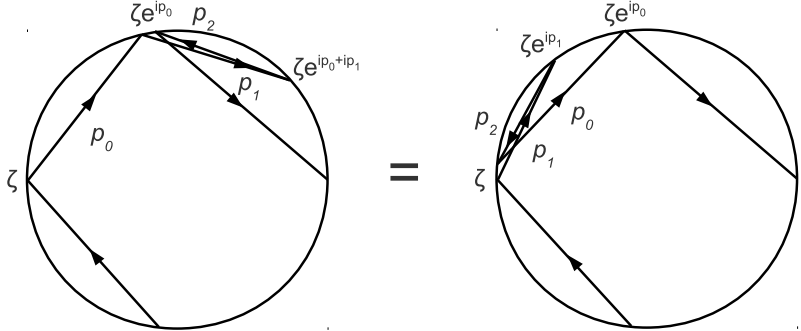,width=10.5cm}\caption[Scattering]
{Scattering of a magnon with a singlet state. The singlet state is
artificially separated a bit for clarity.}
\label{fig_scattering}}
\end{center}\end{figure}

The singlet states are composite states with vanishing total central
charges and zero energy; however, their constituents have non-zero central
charges and energy. These states may be understood as vacuum polarization
states. They can be constructed easily by requiring that they be annihilated
by all symmetry generators \cite{BeisS}. The fundamental singlet
state is\footnote{We denote fundamental states in the bulk by capital letter $A$, 
two-particle bound-states by $B$ and so on. We shall use the lover-case letters 
to denote boundary (bound)states. The bar above the letter denotes anti-particle.}
\begin{align}
\left|1_{12}^{A\bar{A}}\right\rangle \propto & \frac{i\zeta}{\eta_{1}\eta_{2}}\left(1-\frac{x_{1}^{+}}{x_{1}^{-}}\right)\varepsilon^{ab}\omega_{a}^{1}\omega_{b}^{2}+\varepsilon^{\alpha\beta}\theta_{\alpha}^{1}\theta_{\beta}^{2}=-ie^{i\frac{p_{1}}{2}}\,\varepsilon^{ab}\omega_{a}^{1}\omega_{b}^{2}+\varepsilon^{\alpha\beta}\theta_{\alpha}^{1}\theta_{\beta}^{2},\label{singlet_A}
\end{align}
and the relations among the spectral parameters are $x_{2}^{\pm}=1/x_{1}^{\pm}$
($p_{2}=-p_{1}$). Furthermore, there exists a tower of singlet
states\[
\left|1_{12}\right\rangle =\left|1_{12}^{A\bar{A}}\right\rangle +\left|1_{12}^{B\bar{B}}\right\rangle +...\]
emerging from the vacuum as pairs of particle-antiparticle bound-states,
for example
\begin{align}
\left|1_{12}^{B\bar{B}}\right\rangle \propto & \frac{\zeta^{2}}{2\eta_{1}^{2}\eta_{2}^{2}}\frac{\left(y_{1}^{-}-y_{1}^{+}\right)^{2}}{(y_{1}^{-})^{2}}\,\varepsilon^{ac}\varepsilon^{bd}\omega_{a}^{1}\omega_{b}^{1}\omega_{c}^{2}\omega_{d}^{2}\nonumber \\
 & \quad-\frac{i\zeta}{\eta_{1}\eta_{2}}\frac{y_{1}^{-}-y_{1}^{+}}{y_{1}^{-}}\,\varepsilon^{ab}\varepsilon^{\alpha\beta}\omega_{a}^{1}\omega_{b}^{2}\theta_{\alpha}^{1}\theta_{\beta}^{2}+\frac{1}{4}\,\varepsilon^{\alpha\beta}\varepsilon^{\gamma\delta}\theta_{\alpha}^{1}\theta_{\beta}^{1}\theta_{\gamma}^{2}\theta_{\delta}^{2}\nonumber \\
 & =\frac{1}{2}e^{ip_{1}}\,\varepsilon^{ac}\varepsilon^{bd}\omega_{a}^{1}\omega_{b}^{1}\omega_{c}^{2}\omega_{d}^{2}-ie^{i\frac{p_{1}}{2}}\,\varepsilon^{ab}\varepsilon^{\alpha\beta}\omega_{a}^{1}\omega_{b}^{2}\theta_{\alpha}^{1}\theta_{\beta}^{2}+\frac{1}{4}\varepsilon^{\alpha\beta}\varepsilon^{\gamma\delta}\theta_{\alpha}^{1}\theta_{\beta}^{1}\theta_{\gamma}^{2}\theta_{\delta}^{2},\label{singlet_B}
\end{align}
where $y_{2}^{\pm}=1/y_{1}^{\pm}$. These states live in the tensor
product space $V^{*}\otimes V$ of bulk modules, where $V^{*}$ is
the module conjugate to $V$, and are represented as vacuum polarization
bubbles and participate in trivial scattering (see figure \ref{fig_buble}
and figure \ref{fig_scattering}) and play an important role in
crossing symmetry \cite{Janik1,AF2}.


\subsection{Superstrings on the boundary}

We now review  some general concepts for open superstrings in AdS/CFT, linking them to well-known
concepts in the bulk.

\paragraph{The boundary representation.}

Open strings attached to the $Z=0$ giant graviton \cite{HM} and the right factor of
the $Z=0$ $D7$-brane \cite{CY} are two well-known examples of
configurations with a finite-dimensional boundary module respecting
the full bulk symmetry algebra $\g$. The  boundary
$l$-magnon bound-states are described by the following representation
labels \cite{HM}:
\begin{equation}
a_{B}=\sqrt{\frac{g}{2l}}\eta_{B},\quad b_{B}=-\sqrt{\frac{g}{2l}}\frac{i\zeta}{\eta_{B}},
 \quad c_{B}=-\sqrt{\frac{g}{2l}}\frac{\eta_{B}}{\zeta x_{B}},\quad d_{B}=\sqrt{\frac{g}{2l}}\frac{x_{B}}{i\eta_{B}}.\label{abcd_B}
\end{equation}
Unitarity implies $\left|\eta_{B}\right|^{2}=-i\, x_{B}$. We
choose $\eta_{B}={\rm e}^{i\xi}\sqrt{-i\, x_{B}}$ to be the solution
of the unitarity condition, in accordance with the bulk representation.
As in the bulk case, this parametrization of boundary labels
is in the non-local (string) basis. Then by setting $\zeta=1$ one obtains
the local (spin-chain) basis of the $\cal{U}$-deformed
reflection Hopf algebra \eqref{def_Hopf} and \eqref{def_RHopf}.

The multiplet-shortening (mass-shell) condition in terms of $x_B$ is
\begin{equation}
x_{B}+\frac{1}{x_{B}}=i\frac{2l}{g},\label{shortening_b}
\end{equation}
and the eigenvalues of the central charges for the $l$-magnon bound-states
living on the boundary are
\begin{eqnarray}
 & C_{l}=l\, a_{B}b_{B}=-\frac{i}{2}g\,{\rm e}^{2i\xi},\qquad C_{l}^{\dg}=l\, 
 c_{B}d_{B}=\frac{i}{2}g\,{\rm e}^{-2i\xi},\nonumber \\
 & H_{l}=l\left(a_{B}d_{B}+b_{B}c_{B}\right)=\sqrt{l^{2}+g^{2}}.
\end{eqnarray}
As one can see, the spectral parameter $x_{B}$ and the central charge
$H_{l}$ are completely determined by the coupling constant $g$ and the
bound-state number $l$.

We shall use $C(p),\; C^{\dg}(p),\; H(p)$ (bulk)
and $C(q),\; C^{\dg}(q),\; H(q)$ (boundary)
to distinguish the eigenvalues of central charges of bulk and boundary
representations and use the underline ($\ul{\mathcal{O}}$) notation
to distinguish the eigenvalues of bulk charges before and after the
reflection. We shall also use bar ($\bar{\mathcal{O}}$) notation
for the representation of antiparticle states.

\paragraph{The $K$-matrix.}

The $K$-matrix describing the reflection
of bulk states from  boundary states may be represented in
superspace in the same way as the bulk $S$-matrix,
\begin{equation}
\mathcal{K}(p,q)=\sum_{i}k_{i}(p,q)\,\Lambda_{i},
\end{equation}
where $k_{i}(p,q)$ are the reflection coefficients and
the differential operators $\Lambda_{i}$ have the same form as
for the bulk $S$-matrix. Following the same pattern, the reflection
coefficients of the fundamental $K$-matrix may be found by demanding
its invariance (\ref{K_inv}) under the symmetry algebra $\g$
(\ref{g}). Once again, the Lie algebra alone is not enough
to fix uniquely \textit{all} coefficients of the bound-state $K$-matrix
and one additionally needs to use either the boundary Yang-Baxter equation \cite{MR1}
or Yangian symmetry.
\begin{figure}\begin{center}
{\epsfig{file=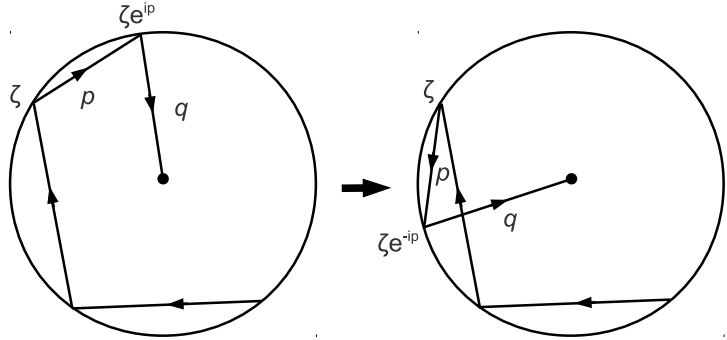,width=9.5cm}\caption[Reflection]
{Reflection from the right boundary of a magnon living on a semi-infinite string
with $\zeta$ being the reference point. The dot in the center of the circle corresponds
to the $Z=0$ giant graviton and the the string ending on it possesses boundary degrees
of freedom.}\label{fig_kmatrix}}\end{center}
\end{figure}

The central charges $\bb{C}$ and $\bb{C}^{\dg}$ of the reflecting 
bulk states are not (on their own) preserved by the reflection, in which
\begin{equation}
\begin{cases}
C(p)=\frac{i}{2}g\left({\rm e}^{ip}-1\right)\zeta, & \hspace{-0.4cm}\\
C^{\dg}(p)=-\frac{i}{2}g\left({\rm e}^{-ip}-1\right)\frac{1}{\zeta}, & \hspace{-0.4cm}\end{cases}\rightarrow\begin{cases}
{C}(-p)=\frac{i}{2}g\left({\rm e}^{-ip}-1\right)\zeta,\\
{C}^{\dg}(-p)=-\frac{i}{2}g\left({\rm e}^{ip}-1\right)\frac{1}{\zeta},\end{cases}
\end{equation}
but they {\em are} preserved by the reflection from a boundary
with degrees of freedom, in the sense that the total central charges
are conserved
\begin{eqnarray}
C(p)+\mathcal{U}^{+2}C(q) & = & \frac{i}{2}g\left({\rm e}^{ip}-1\right)\zeta-\frac{i}{2}g\zeta{\rm e}^{ip}=-\frac{i}{2}g\zeta\equiv\ul{C}(p)+\mathcal{U}^{-2}C(q),\nonumber \\
C^{\dg}(p)+\mathcal{U}^{-2}C^{\dg}(q) & = & -\frac{i}{2}g\left({\rm e}^{-ip}-1\right)\frac{1}{\zeta}+\frac{i}{2}g\frac{1}{\zeta}{\rm e}^{-ip}=\frac{ig}{2\zeta}\equiv\ul{C}^{\dg}(p)+\mathcal{U}^{+2}C^{\dg}(q),
\end{eqnarray}
as required from the considerations of the reflection algebra \eqref{cons_center}.

\paragraph{The spectrum and the special points.}

The spectrum of boundary excitations is very much the same as the spectrum
of bulk states; it includes fundamental, bound-states and singlet
(mirror) states \cite{HM,CY,MR1,LP,ABR,BP}

\begin{figure}\begin{center}
\epsfig{file=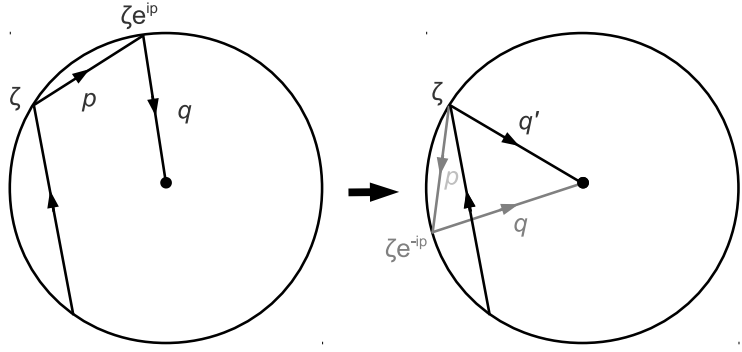,width=9.7cm}\caption[Bound-state]
{Construction of a two-particle boundary bound-state appearing at the
pole $x^{-}=x_{B}$ of the $K$-matrix. The spectral parameter of the
emerging boundary bound-state is $y_{B} = x_{B} e^{i p} = x_{B}
\frac{x^{+}}{x^{-}} = x_{+}$.}\label{fig_bbs}
\end{center}\end{figure}

Boundary bound-states appear as poles $x^{-}=x_{B}$ in the $K$-matrix
projecting it to the totally symmetric part of the tensor product $V(p)\otimes V_{B}(q)$
of bulk and boundary modules signaling the presence of the tower
of multi-particle boundary bound-states. The boundary spectral
parameter $x_{B}$ is purely imaginary, hence only the states with
real momentum $p$ (physical states) may be absorbed by the boundary to form a bound-state.
The new bound state has spectral parameter set to $y_{B}=x^{+}$ (see
figure \ref{fig_bbs}).

The boundary singlet states, like the bulk singlet states,
are composite states with vanishing total central charges and zero
energy, while the constituent states have non-zero central charges
and energies. These states may be formally understood as the excitations 
of the corresponding $D$-branes.\footnote{One must not forger that
these are still `very long' spin chains with a very large number of bulk
vacuum fields $Z$ but without any bulk excitations.}
Boundary singlet states are constructed in the same way as bulk singlet
states, by requiring their annihilation by all symmetry generators;
one finds a tower of boundary singlet (bound)states
\begin{equation}
\left|1_{12}\right\rangle _{B}=\left|1_{12}^{\bar{a}a}\right\rangle _{B}+\left|1_{12}^{\bar{b}b}\right\rangle _{B}+...,\label{singlet_boundary}
\end{equation}
where
\begin{align}
\left|1_{12}^{\bar{a}a}\right\rangle _{B}\propto & \frac{i\zeta}{\eta_{1}\eta_{2}}\varepsilon^{ab}\omega_{a}^{1}\omega_{b}^{2}+\varepsilon^{\alpha\beta}\theta_{\alpha}^{1}\theta_{\beta}^{2}\nonumber \\
 & =-i\,\varepsilon^{ab}\omega_{a}^{1}\omega_{b}^{2}+\varepsilon^{\alpha\beta}\theta_{\alpha}^{1}\theta_{\beta}^{2},\label{singlet_a}\\
\left|1_{12}^{\bar{b}b}\right\rangle _{B}\propto & \frac{\zeta^{2}}{2\eta_{1}^{2}\eta_{2}^{2}}\varepsilon^{ac}\varepsilon^{bd}\omega_{a}^{1}\omega_{b}^{1}\omega_{c}^{2}\omega_{d}^{2}-\frac{i\zeta}{\eta_{1}\eta_{2}}\varepsilon^{ab}\varepsilon^{\alpha\beta}\omega_{a}^{1}\omega_{b}^{2}\theta_{\alpha}^{1}\theta_{\beta}^{2}+\frac{1}{4}\varepsilon^{\alpha\beta}\varepsilon^{\gamma\delta}\theta_{\alpha}^{1}\theta_{\beta}^{1}\theta_{\gamma}^{2}\theta_{\delta}^{2}\nonumber \\
 & =\frac{1}{2}\varepsilon^{ac}\varepsilon^{bd}\omega_{a}^{1}\omega_{b}^{1}\omega_{c}^{2}\omega_{d}^{2}+i\,\varepsilon^{ab}\varepsilon^{\alpha\beta}\omega_{a}^{1}\omega_{b}^{2}\theta_{\alpha}^{1}\theta_{\beta}^{2}+\frac{1}{4}\varepsilon^{\alpha\beta}\varepsilon^{\gamma\delta}\theta_{\alpha}^{1}\theta_{\beta}^{1}\theta_{\gamma}^{2}\theta_{\delta}^{2},\label{singlet_b}
\end{align}
plus the pairs of higher-order bound-states. The boundary singlet
states live in the tensor product space $V_{B}(\bar{q})\otimes V_{B}(q)$
where $V_{B}(\bar{q})$ is the conjugate boundary module $V_{B}^{*}$
with respect to $V_{B}$. The boundary phases are $\zeta^{\bar{q}}=-\zeta$
and $\zeta^{q}=\zeta$, and the spectral parameters are related as
$x_{B}^{\bar{q}}=1/x_{B}^{q}$, where we have added extra indices
$\bar{q}$ and $q$ to distinguish parameters of the left and right 
boundaries respectively.

\begin{figure}\begin{center}
{\epsfig{file=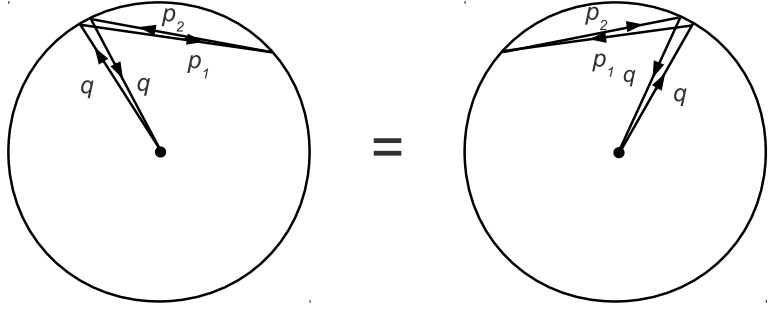,width=10cm}\caption[Excitation]
{Open string as an excitation of a $Z=0$ giant graviton or $D7$-brane.}
\label{fig_excitation}}
\end{center}\end{figure}

The bulk and boundary singlet states together form an open string
configuration which may be considered as an excitation of the $Z=0$
giant graviton or (in the right factor of) the $Z=0$ $D7$-brane (see figure \ref{fig_excitation}).
This configuration clearly has vanishing total central charges and
is annihilated by all symmetry generators. We believe this state might
be useful in calculating the dressing factor of the $K$-matrix.


\subsection{Special points and automorphisms}\label{secAut}

The bulk mass-shell condition
\begin{equation}
x^{+}+\frac{1}{x^{+}}-x^{-}-\frac{1}{x^{-}}=i\frac{2l}{g},
\end{equation}
admits three automorphism maps $x^{\pm}\mapsto1/x^{\pm}$, $x^{\pm}\mapsto-x^{\mp}$
and $x^{\pm}\mapsto-1/x^{\mp}$, but actually only two maps are independent: 
the composition of any two gives the third. We shall use
the first two maps, corresponding to the particle-antiparticle map
and the reflection map.

The boundary mass-shell condition
\begin{equation}
x_{B}+\frac{1}{x_{B}}=i\frac{2l}{g},
\end{equation}
is much less rich in automorphisms and admits only one automorphism,
$x_{B}\mapsto1/x_{B}$, which corresponds to the particle-antiparticle
map.

\paragraph{Particle-antiparticle map.}

The automorphism $x^{\pm}\mapsto1/x^{\pm}$
maps the central charges and momentum to the opposite values
$\left(H,C,C^{\dg},p\right)\mapsto\left(\bar{H},\bar{C},
\bar{C}^{\dg},\bar{p}\right)$,\footnote{Here we use the notation that $\bar{H}=-H$, $\bar{p}=-p$
and $\bar{C}$, $\bar{C}^\dg$ are the opposite values as described in \cite{AF3}.}
i.e.\ it maps the state on the positive energy branch of the dispersion
relation into the state on the negative branch and vice versa; this
is the analogue of the crossing-symmetry transformation in two-dimensional
relativistic field theories.

Let the states (of positive energy) transform in some module $V$
of the symmetry algebra $\g$. Then the antiparticle states
(of negative energy) transform in the conjugate module $V^{*}$
of $\g$, which is equivalent to $V$ up to the isomorphism
$\rho$
\begin{equation}
V=\mathcal{C}^{-1}\rho\left(V^{*}\right)\mathcal{C},\label{V/V*}
\end{equation}
where $\mathcal{C}$ is an particle-antiparticle flavor-intertwining
matrix. Let $\pi(\g)$ be the matrix representation of the
algebra on the positive energy (particle) states and $\bar{\pi}(\g)$
on the corresponding negative energy (antiparticle) states.
Then the representation labels (\ref{abcd})
of antiparticle and particle states are related by
\begin{equation}
\bar{a}(p)=-\frac{i}{x^{+}}\, a(p),\;\bar{b}(p)=-ix^{-}\, b(p),\;\bar{c}(p)=-ix^{+}\, c(p),\;\bar{d}=-\frac{i}{x^{-}}\, d(p),
\end{equation}
and the map $\rho$ explicitly reads as\footnote{This map was first
constructed in \cite{Janik1}; as a nice review we recommend \cite{AF3}.}
\begin{equation}
\rho:\bar{\pi}(\g)\mapsto-\bar{\pi}\left(\wh{\rho}\circ\g\right)^{st},\label{q_map}
\end{equation}
where $st$ is the super-transpose and $\wh{\rho}$ is a $U(1)$-automorphism
which, combined with the minus sign, is equivalent to the antipode map
$S$ of the algebra, i.e.\ $\rho:\bar{\pi}\left(\g\right)\mapsto\bar{\pi}\left(S\circ\g\right)^{st}$;
hence (\ref{V/V*}) explicitly reads as
\begin{equation}
\pi(\g)=\mathcal{C}^{-1}\,\bar{\pi}\left(S\circ\g\right)^{st}\,\mathcal{C}.\label{Pi/Pi^bar}
\end{equation}
It is quite straightforward to check that the relation above fixes
$\mathcal{C}$ up to an overall factor to be
\begin{equation}
\mathcal{C}=\left(\begin{array}{cc}
\sigma_{2} & 0\\
0 & i\sigma_{2}\end{array}\right).
\end{equation}
The same result can be easily obtained merely by looking at the singlet
state (\ref{singlet_A}).

Let us examine what happens when the symmetry is lifted to the Yangian
symmetry. Let $\pi(ev_{v}(\mbox{Y}(\g)))$ be the evaluation
representation of the Yangian algebra of the positive energy (particle)
states and $\bar{\pi}(ev_{v}(\mbox{Y}(\g)))$ be the evaluation
representation of the corresponding negative energy (antiparticle)
states. The spectral parameter $v$ of the evaluation map (\ref{vJ_ansatz})
is identified with the rapidity $u$ of the magnon \cite{BeisY}.
The rapidity $u$ is invariant under the map $x^{\pm}\mapsto1/x^{\pm}$ ---
that is, it is mapped to itself $u\mapsto u$ by the particle-antiparticle
map. Hence it is easy to see that (\ref{Pi/Pi^bar_B}) trivially
lifts to
\begin{equation}
\pi(ev_{u}(\mbox{Y}(\g)))=\mathcal{C}^{-1}\,\bar{\pi}(ev_{u}(\mbox{Y}(S\circ\g)))^{st}\,\mathcal{C},\label{Y_Pi/Pi^bar}
\end{equation}
which says that Yangian charges behave in the same way as the
algebra generators under the particle-antiparticle map.

The automorphism $x_{B}\mapsto1/x_{B}$ of the boundary mass-shell
condition maps the energy of the boundary state to the opposite branch
of the dispersion relation, but keeps the eigenvalues of the boundary central
charges $C$ and $C^{\dg}$ unchanged. Thus to use this map to construct the
particle-antiparticle map one needs to consider a composition of the
map $x_{B}\mapsto1/x_{B}$ together with the map $\zeta\mapsto-\zeta$
which sends the central charges $C$ and $C^{\dg}$ to their opposite
values and may be interpreted as an interchange of the left and right
boundaries. This construction of the particle-antiparticle map is
in agreement with the boundary singlet state (\ref{singlet_boundary})
which is required to be invariant under it.

Following the construction of the particle-antiparticle maps in the
bulk, the boundary states of positive energy transform in a boundary
module $V_{B}$ and the states of the negative energy transform in
the conjugate module $V_{B}^{*}$, which is equivalent to $V_{B}$
up to the isomorphism $\rho_{B}$ such that
\begin{equation}
V_{B}=\mathcal{C}^{-1}\rho_{B}\left(V_{B}^{*}\right)\mathcal{C},\label{VB/VB*}
\end{equation}
where $\mathcal{C}$ is a particle-antiparticle flavor-intertwining
matrix and is clearly the same for bulk and boundary representations.
As for the bulk, let $\pi_{B}(\h)$ be the matrix representation of the boundary
algebra on the positive-energy states and $\bar{\pi}_{B}(\h)$
on the corresponding negative-energy states. The boundary representation labels (\ref{abcd_B})
of the antiparticles and particles are related by
\begin{equation}
\bar{a}(q)=-\frac{i}{x_{B}}\, a(q),\;\bar{b}(q)=-ix_{B}\, b(q), \;\bar{c}(q)=ix_{B}\, c(q),\;\bar{d}=-\frac{i}{x_{B}}\, d(q).
\end{equation}
Then the map (\ref{q_map}) in the case of the boundary representation becomes
\begin{equation}
\rho_{B}:\bar{\pi}_{B}(\h)\mapsto-\bar{\pi}_{B}\left(\wh{\rho}_{B}\circ\h\right)^{st},
\end{equation}
leading to the boundary partner of the relation (\ref{Pi/Pi^bar})
\begin{equation}
\pi_{B}(\h)=\mathcal{C}^{-1}\,\bar{\pi}_{B}\left(S\circ\h\right)^{st}\,\mathcal{C},\label{Pi/Pi^bar_B}
\end{equation}
where the boundary representation of the braiding factor is trivial
\begin{equation}
\pi_{B}(\mathcal{U})=\bar{\pi}_{B}(\mathcal{U})=1,
\end{equation}
implying that the boundary representation of the antipode map $S$
is
\begin{equation}
\pi_{B}(S(1))=\pi_{B}(S(\mathcal{U}))=1,\quad\pi_{B}(S(\bb{J}^{A}))=-\pi_{B}(\bb{J}^{A}).
\end{equation}

We conclude by constructing the representation of the boundary Yangian
by defining $\pi_{B}(ev_{w}(\mbox{Y}(\g)))$ to be the evaluation
representation of the positive energy (particle) states and
$\bar{\pi}_{B}(ev_{w}(\mbox{Y}(\g)))$
to be that of the corresponding negative energy
(antiparticle) states. We shall explicitly define the on-shell realization
of the spectral parameter $w$ of the evaluation map of the boundary
representation in section 4.5. For now we only require it to be invariant
under the particle-antiparticle maps in accordance with the bulk representation.
Then the relation (\ref{Pi/Pi^bar_B}) is being lifted to
\begin{equation}
\pi_{B}(ev_{w}(\mbox{Y}(\g)))=\mathcal{C}^{-1}\,\bar{\pi}_{B}(ev_{w}(\mbox{Y}(S\circ\g)))^{st}\,\mathcal{C}.
\end{equation}

\paragraph{Reflection map.}

The automorphism $x^{\pm}\mapsto-x^{\mp}$
maps the momentum of the state to the opposite value $p\mapsto-p$ while
preserving the energy $H\mapsto H$. Reflection is an involutive
map, as it keeps the states on the same branch of the dispersion relation,
---that is, it is an automorphism of $V$ (or $V^{*}$), and may be considered
as the analogue of parity symmetry in two-dimensional relativistic
field theories. Let us consider the reflection of  states transforming
in the module $V$. Let $\pi(\g)$ be the matrix
representation of the algebra on incoming states and $\ul{\pi}(\g)$
 on the reflected states.
The representation labels (\ref{abcd}) are related as
\begin{equation}
\ul{a}(p)=\sqrt{\frac{x^{-}}{x^{+}}}\, a(p),\;\ul{b}(p)=-\sqrt{\frac{x^{-}}{x^{+}}}\, b(p),\;\ul{c}(p)=-\sqrt{\frac{x^{+}}{x^{-}}}\,c(p),\;\ul{d}(p)=\sqrt{\frac{x^{+}}{x^{-}}}\, d(p),\label{reptilde/rep}
\end{equation}
while the central charges $C$ and $C^{\dg}$ become
\begin{equation}
\ul{C}(p)=-\frac{x^{-}}{x^{+}}C(p),\qquad\ul{C}^{\dg}(p)=-\frac{x^{+}}{x^{-}}C^{\dg}(p),\label{CC_ref}
\end{equation}
where the underline notation once again denotes the eigenvalues of the
charges after the reflection. Then the two representations $\pi(\g)$
and $\ul{\pi}(\g)$ are equivalent to each other up to
the isomorphism $\kappa$
\begin{equation}
\pi(\g)=\mathcal{P}^{-1}\,\ul{\pi}(\kappa\circ\g)\,\mathcal{P},\label{ref_relation}
\end{equation}
where $\mathcal{P}$ is the parity transformation matrix and is of
 diagonal form. The relations (\ref{reptilde/rep}) clearly indicate
that the reflection may be realized as a $U(1)$-automorphism
$\kappa$ of the algebra. By solving (\ref{ref_relation})
we find that ${\kappa}$ is a map under which the representation
parameters undergo a right shift
\begin{equation}
\kappa:\left(\begin{array}{cc}
a & b\\
c & d\end{array}\right)\mapsto\left(\begin{array}{cc}
-i\,\mathcal{U}^{-1} & 0\\
0 & i\,\mathcal{U}^{+1}\end{array}\right)\left(\begin{array}{cc}
a & b\\
c & d\end{array}\right).
\end{equation}
Its pull-back map acts on the supersymmetry charges as
\begin{equation}
\kappa:\bigl(\bb{Q}_{\alpha}^{\; a},\bb{G}_{b}^{\;\beta}\bigr)\mapsto\bigl(-i\,\mathcal{U}^{-1}\,\bb{Q}_{\alpha}^{\; a},i\,\mathcal{U}^{+1}\,\bb{G}_{b}^{\;\beta}\bigr),
\end{equation}
and on the central charges $\bb{C}$ and $\bb{C}^{\dg}$ as
\begin{equation}
\kappa:\bigl(\bb{C},\bb{C}^{\dg}\bigr)\mapsto\bigl(-\mathcal{U}^{-2}\,\bb{C},-\mathcal{U}^{+2}\,\bb{C}^{\dg}\bigr),
\end{equation}
while it acts trivially on all other generators. The reflection map fixes
the parity transformation matrix to be
\begin{equation}
\mathcal{P}=\left(\begin{array}{cc}
\mathbbm{1}_{2} & 0\\
0 & -i\mathbbm{1}_{2}\end{array}\right).
\end{equation}

We proceed by lifting the representations of the reflection algebra to the Yangian.
Let $\pi(ev_{u}(\mbox{Y}(\g)))$ be the evaluation representation
of the Yangian algebra on the incoming states and $\ul{\pi}(ev_{u}(\mbox{Y}(\g)))$
on the reflected states. The reflection
map $x^{\pm}\mapsto-x^{\mp}$ results in $u\mapsto-u$
for the rapidity of the state, leading to a minus sign appearing in
the evaluation map (\ref{ev_map}). However, the parity map \eqref{ParOp}
reverses the sign of rapidity again thus the algebra representation
relation (\ref{ref_relation}) is naturally lifted
to the level of the evaluation representation
\begin{equation}
\pi(ev_{u}(\mbox{Y}(\g)))=\mathcal{P}^{-1}\,\ul{\pi}(ev_{u}(\mbox{Y}(\kappa\circ\g)))\,\mathcal{P}^{-1}.\label{Y_Pi/Pi^tilde}
\end{equation}


\subsection{The representation of the Yangian symmetry}

The evaluation representation of the bulk Yangian symmetry was first
considered in \cite{BeisY}. We shall extend this to include the boundary representation
and the constraints arising from the reflection algebra.

\paragraph{Bulk case.}

The evaluation representation is constructed with the help
of the evaluation map ansatz \eqref{vJ_ansatz} that we have considered
in section 2.2,
\begin{equation}\label{evansatz}
\wh{\bb{J}}\left|v\right\rangle =\gamma\,(v+v_{0})\,\bb{J}\left|v\right\rangle .
\end{equation}
The representation parameters $\gamma$ and $v$ may be defined by 
considering the co-commutativity of the central charges of the algebra.

The deformed central charges (\ref{HCC_twist}) in the evaluation
representation \eqref{evansatz} of the bulk module become
\begin{align}
\wh{\bb{C}}' & =\gamma\left(v+v_{0}\right)\bb{C}+\frac{1}{2}\bb{H}\left(\bb{C}-2\alpha\right)=\beta\, v_{C}\,\bb{C},\nonumber \\
\wh{\bb{C}}^{\dg\prime} & =\gamma\left(v+v_{0}\right)\bb{C}^{\dg}-\frac{1}{2}\bb{H}\,(\bb{C}^{\dg}-2\alpha^{\dg})=\beta\, v_{C^\dg}\,\bb{C}^{\dg},\label{HCC_deformed_onshell}
\end{align}
where we have used constraints (\ref{CC_twist_ansatz}) and 
(\ref{H_twist_ansatz}). Assuming that all universal constant are equal $v_C = v_C^\dg=v_0$, these equations 
have a family of solutions of the form
\begin{equation}
v=\frac{ig}{2\gamma}u+\frac{\beta}{\gamma}v_{0}-v_{0},
\end{equation}
where $u$ is the rapidity \eqref{rapidity} of the corresponding magnon and the parameters $\beta$
and $\gamma$ are left unconstrained. The most natural choice is
\begin{equation}
\beta=\gamma=i\frac{g}{2},\label{beta_gamma}
\end{equation}
giving a simple, sensible solution $v=u$ while keeping $v_0$ unconstrained.

\paragraph{Boundary case.}

We now proceed by considering the evaluation representation of $\rm Y(\g,\g)$ on the boundary module.
We assume that a similar ansatz to \eqref{ev2_map} holds for the boundary module
\begin{equation}\label{ev2_mapB}
\dwh{\bb{J}}\left|w\right\rangle =\gamma_B^2\,(w+w_{0})^2\,\bb{J}\left|w\right\rangle .
\end{equation}
where $\left|w\right\rangle \in V_B(w)$ is a vector of the boundary module. 
Then consistency with the evaluation representation for the bulk module 
requires $w_0 = v_0$ and $\gamma_B = \gamma$, leading to the following 
evaluation representation of the charges \eqref{C2nc}
\begin{align}
\dwh{\bb{C}}\,'\otimes1 &= -\frac{1}{4}g^2 (u+v_0)^2\, \bb{C}\otimes1,
& 1\otimes\dwh{\bb{C}}\,' &= 1\otimes\Bigl(-\frac{1}{4}g^2 (w+v_0)^2\Bigr)\bb{C}, \nonumber\\
\dwh{\bb{C}}\,^{\dg\prime}\otimes1 &= -\frac{1}{4}g^2 (u+v_0)^2\, \bb{C}^{\dg}\otimes1,
& 1\otimes\dwh{\bb{C}}\,^{\dg\prime} &= 1\otimes\Bigl(-\frac{1}{4}g^2 (w+v_0)^2\Bigr) \bb{C}^\dg.
\end{align}
Further, requiring co-conservation constraint \eqref{C2cons} to hold, 
we find $w^2 = -l^2/g^2$ and also $v_0=0$. Then by choosing the positive root we 
define the spectral parameter associated to the boundary module to be
\begin{align}\label{brapidity}
w=\frac{il}{g}.
\end{align}
Let us discuss the origin of this boundary spectral parameter. It is
easy to observe that boundary labels \eqref{abcd_B} and mass-shell
condition \eqref{shortening_b} may be obtained from the bulk ones
\eqref{abcd} and \eqref{shortening} by a simple map $x^\pm\mapsto\pm \xb,\;l
\mapsto l/2$. This relation says that a $l$-magnon bound-state on the
boundary may be viewed as a $2l$-magnon bound-state in the bulk of
maximal momentum $p=\pi$ \cite{ABR}. Thus applying this map to the bulk
rapidity \eqref{rapidity}
\begin{align}
 u=\xp+\frac{1}{\xp}-\frac{il}{g}\;\longmapsto\;\xb
+\frac{1}{\xb}-\frac{il}{g}=2\frac{il}{g}-\frac{il}{g}=w,
\end{align}
we get the boundary one \eqref{brapidity}, as required.

It is interesting to compare this case with the reflection from the
`vertical' $D5$-brane considered in \cite{CY,CRY,MR3}, which also has
boundary degrees of freedom attached to the boundary. The boundary
labels and mass-shell condition of the `vertical' $D5$-brane may be
obtained from those of $Z=0$ giant graviton (and the right factor of
$Z=0$ $D7$ brane) by replacing $l\mapsto l/2$. Thus applying the map $x^\pm
\mapsto\pm x_B$ to bulk labels one recovers exactly the boundary labels of
the `vertical' $D5$-brane, implying that the $l$-magnon bound-state on
the boundary is equivalent to a $l$-magnon bound-state in the bulk with
momentum $p=\pi$. Furthermore, this results in the boundary spectral
parameter $w_{D5}=0$ being zero
\begin{align}
 u=\xp+\frac{1}{\xp}-\frac{il}{g}\;\underset{D5}{\longmapsto}\;\xb
+\frac{1}{\xb}-\frac{il}{g}=\frac{il}{g}-\frac{il}{g}=0,
\end{align}
as was observed in \cite{MR3}.

With the help of the evaluation representation \eqref{ev2_mapB}, 
\eqref{brapidity} of the boundary Yangian $\rm Y(\g,\g)$ we have 
checked explicitly that the twisted charges \eqref{J2Z} are the 
symmetries of the fundamental reflection matrix of the $Z=0$ giant 
graviton \cite{HM} and the right factor of the $Z=0$ $D7$ 
brane \cite{CY,MR1}. We have also checked that these charges 
define uniquely the reflection matrix of bulk two-magnon bound-states 
reflecting from boundary two-magnon bound-states and is consistent 
with the boundary Yang-Baxter equation.\footnote{Interestingly, 
there is also an apparently accidental symmetry of the fundamental 
reflection matrix satisfying the invariance condition
$\mathcal{K}\,\Delta(\wh{\bb{J}})-\Delta^{\rfl\;\prime}(\wh{\bb{J}})\,\mathcal{K}=0$, where
$\Delta\wh{\bb{J}}$ is the usual coproduct of level-1 charges, but $\Delta^{\rfl\;\prime}$ 
is a `modified' reflected coproduct
\begin{align}
\Delta^{\rfl\;\prime}(\wh{\bb{J}}^{A})=-\wt{\wh{\bb{J}}}{}^{A}\otimes1+\mathcal{U}^{-\left[A\right]}\otimes\wh{\bb{J}}^{A}-f_{\; BC}^{A}\,\mathcal{U}^{-\left[C\right]}\,\wt{\bb{J}}^{B}\otimes\bb{J}^{C},
\end{align}
and also the boundary is required to carry a `strange' spectral parameter
\begin{align}
w=\frac{1}{2}\left(x_{B}-\frac{1}{x_{B}}\right)\frac{x^{-}-x^{+}}{x^{-}+x^{+}}+
\frac{1+x^{-}x^{+}}{x^{-}+x^{+}}.
\end{align}
This symmetry was also observed by T. Matsumoto and R. Nepomechie 
(private discussions). However, this is not a symmetry of the bound-state 
reflection matrices, as it leads to a unique solution of the invariance 
condition which is not consistent with the boundary Yang-Baxter equation.
}


\section{Conclusions}

In this paper we have presented the `Heisenberg picture' of
the reflection algebra and constructed the twisted boundary Yangian
symmetry of the AdS/CFT superstring ending on a boundary with degrees
of freedom and preserving the full bulk Lie algebra. The two known
boundaries of this type are the $Z=0$ giant graviton ($D3$-brane) and
the right factor of the $Z=0$ $D7$-brane.

The corresponding Yangian symmetry $\rm Y(\g,\g)$ is a fixed-point co-ideal
subalgebra of the bulk Yangian $\rm Y(\g)$ generated by level-0 and twisted
level-2 generators which may be obtained by considering quadratic combinations
of level-1 Yangian charges and adding appropriate deformation terms (`twists') that ensure
the co-ideal property, which here requires that the co-products of twisted level-2 charges
include no level-1 charges acting on the boundary module.

We have explicitly constructed the reflection automorphism of the
AdS/CFT superstring symmetry algebra and presented the `reflection
Hopf algebra', which we showed to be a natural extension of the usual
Hopf algebra structure to accommodate reflection. The notion of reflected
coproducts allows us to consider the boundary representation in the `spin-chain'
basis, in which boundary representation labels are independent of the
phase of bulk magnons, and thus are of `local' form. Furthermore, the
reflection Hopf algebra structure allowed us to construct the boundary
Yangian and its evaluation representation in a manner very similar
to that for the bulk case \cite{BeisY}.

We have also considered the spectrum of states of the open strings.
We have shown that the boundary spectrum includes not only bound-states
but also singlet states very similar to those of the bulk excitation spectrum.
Furthermore, we have shown
that there is a tower of bulk and boundary singlet states, and
have explicitly constructed such states
from two-magnon bound-states. Such states could form an open spin-chain
configuration that might perhaps be understood as a $D$-brane excitation (figure
4). We certainly expect this configuration to be useful for finding the dressing
phase of the reflection matrices.

The most important remaining open question regarding superstrings ending on $D7$-branes
is to determine the Yangian symmetry of the left factor of the $Z=0$ $D7$-brane.
The other interesting question is how the twisted Yangians of $D3$-, $D5$- and $D7$-branes
could be obtained from the quantum affine algebra recently constructed in \cite{BGM}, and what are the corresponding twisted
affine algebras.

\textbf{Acknowledgments}:
We thank Charles Young and Evgeni Sklyanin for valuable discussions, and Alessandro Torrielli for many
fruitful conversations and for reading the manuscript.
V.R.\ thanks Marius de Leeuw for sharing his \textit{Mathematica}
code for calculating bound-state $S$-matrices, and which we used to double-check our results,
and also Takuya Matsumoto and Rafael Nepomechie for useful discussions, and
the Galileo Galilei Institute for Theoretical Physics for hospitality
while part of this work was done.
The authors also thank the UK EPSRC for funding under grant EP/H000054/1.


\appendix

\section{The co-products of Y$(\g)$}\label{appA}

The co-products of the Yangian charges of the cetrally extended 
$\mathfrak{psu}(2|2)\ltimes\bb{R}^3$ are \cite{BeisY}
\begin{align}
\Delta\wh{\bb{R}}_{a}^{\enskip b} & =\wh{\bb{R}}_{a}^{\enskip b}\otimes1+1\otimes\wh{\bb{R}}_{a}^{\enskip b}+\frac{1}{2}\bb{R}_{a}^{\enskip c}\otimes\bb{R}_{c}^{\enskip b}-\frac{1}{2}\bb{R}_{c}^{\enskip b}\otimes\bb{R}_{a}^{\enskip c}\nonumber \\
 & \qquad-\frac{1}{2}\bb{G}_{a}^{\enskip\gamma}\,\mathcal{U}^{+1}\otimes\bb{Q}_{\gamma}^{\enskip b}-\frac{1}{2}\bb{Q}_{\gamma}^{\enskip b}\,\mathcal{U}^{-1}\otimes\bb{G}_{a}^{\enskip\gamma}\nonumber \\
 & \qquad+\frac{1}{4}\delta_{a}^{b}\bb{G}_{c}^{\enskip\gamma}\,\mathcal{U}^{+1}\otimes\bb{Q}_{\gamma}^{\enskip c}+\frac{1}{4}\delta_{a}^{b}\bb{Q}_{\gamma}^{\enskip c}\,\mathcal{U}^{-1}\otimes\bb{G}_{c}^{\enskip\gamma},\nonumber \\
\Delta\wh{\bb{L}}_{\alpha}^{\enskip\beta} & =\wh{\bb{L}}_{\alpha}^{\enskip\beta}\otimes1+1\otimes\wh{\bb{L}}_{\alpha}^{\enskip\beta}-\frac{1}{2}\bb{L}_{\alpha}^{\enskip\gamma}\otimes\bb{L}_{\gamma}^{\enskip\beta}+\frac{1}{2}\bb{L}_{\gamma}^{\enskip\beta}\otimes\bb{L}_{\alpha}^{\enskip\gamma}\nonumber \\
 & \qquad+\frac{1}{2}\bb{G}_{c}^{\enskip\beta}\,\mathcal{U}^{+1}\otimes\bb{Q}_{\alpha}^{\enskip c}+\frac{1}{2}\bb{Q}_{\alpha}^{\enskip c}\,\mathcal{U}^{-1}\otimes\bb{G}_{c}^{\enskip\beta}\nonumber \\
 & \qquad-\frac{1}{4}\delta_{\alpha}^{\beta}\,\bb{G}_{c}^{\enskip\gamma}\,\mathcal{U}^{+1}\otimes\bb{Q}_{\gamma}^{\enskip c}-\frac{1}{4}\delta_{\alpha}^{\beta}\,\bb{Q}_{\gamma}^{\enskip c}\,\mathcal{U}^{-1}\otimes\bb{G}_{c}^{\enskip\gamma},\nonumber \\
\Delta\wh{\bb{Q}}_{\alpha}^{\enskip a} & =\wh{\bb{Q}}_{\alpha}^{\enskip a}\otimes1+\mathcal{U}^{+1}\otimes\wh{\bb{Q}}_{\alpha}^{\enskip a}+\frac{1}{2}\bb{Q}_{\alpha}^{\enskip c}\otimes\bb{R}_{c}^{\enskip a}-\frac{1}{2}\bb{R}_{c}^{\enskip a}\,\mathcal{U}^{+1}\otimes\bb{Q}_{\alpha}^{\enskip c}\nonumber \\
 & \qquad+\frac{1}{2}\bb{Q}_{\gamma}^{\enskip a}\otimes\bb{L}_{\alpha}^{\enskip\gamma}-\frac{1}{2}\bb{L}_{\alpha}^{\enskip\gamma}\,\mathcal{U}^{+1}\otimes\bb{Q}_{\gamma}^{\enskip a}+\frac{1}{4}\bb{Q}_{\alpha}^{\enskip a}\otimes\bb{H}-\frac{1}{4}\bb{H}\,\mathcal{U}^{+1}\otimes\bb{Q}_{\alpha}^{\enskip a}\nonumber \\
 & \qquad+\frac{1}{2}\varepsilon_{\alpha\gamma}\varepsilon^{ac}\,\bb{C}\,\mathcal{U}^{-1}\otimes\bb{G}_{c}^{\enskip\gamma}-\frac{1}{2}\varepsilon_{\alpha\gamma}\varepsilon^{ac}\,\bb{G}_{c}^{\enskip\gamma}\otimes\bb{C},\nonumber \\
\Delta\wh{\bb{G}}_{a}^{\enskip\alpha} & =\wh{\bb{G}}_{a}^{\enskip\alpha}\otimes1+\mathcal{U}^{-1}\otimes\wh{\bb{G}}_{a}^{\enskip\alpha}-\frac{1}{2}\bb{G}_{c}^{\enskip\alpha}\otimes\bb{R}_{a}^{\enskip c}+\frac{1}{2}\bb{R}_{a}^{\enskip c}\,\mathcal{U}^{-1}\otimes\bb{G}_{c}^{\enskip\alpha}\nonumber \\
 & \qquad-\frac{1}{2}\bb{G}_{a}^{\enskip\gamma}\otimes\bb{L}_{\gamma}^{\enskip\alpha}+\frac{1}{2}\bb{L}_{\gamma}^{\enskip\alpha}\,\mathcal{U}^{-1}\otimes\bb{G}_{a}^{\enskip\gamma}-\frac{1}{4}\bb{G}_{a}^{\enskip\alpha}\otimes\bb{H}+\frac{1}{4}\bb{H}\,\mathcal{U}^{-1}\otimes\bb{G}_{a}^{\enskip\alpha}\nonumber \\
 & \qquad-\frac{1}{2}\varepsilon_{ac}\varepsilon^{\alpha\gamma}\,\bb{C}^{\dg}\,\mathcal{U}^{+1}\otimes\bb{Q}_{\gamma}^{\enskip c}+\frac{1}{2}\varepsilon_{ac}\varepsilon^{\alpha\gamma}\,\bb{Q}_{\gamma}^{\enskip c}\otimes\bb{C}^{\dg},\nonumber \\
\Delta\wh{\bb{H}} & =\wh{\bb{H}}\otimes1+1\otimes\wh{\bb{H}}+\bb{C}\,\mathcal{U}^{-2}\otimes\bb{C}^{\dg}-\bb{C}^{\dg}\,\mathcal{U}^{+2}\otimes\bb{C},\nonumber \\
\Delta\wh{\bb{C}} & =\wh{\bb{C}}\otimes1+\mathcal{U}^{+2}\otimes\wh{\bb{C}}-\frac{1}{2}\bb{H}\,\mathcal{U}^{+2}\otimes\bb{C}+\frac{1}{2}\bb{C}\otimes\bb{H},\nonumber \\
\Delta\wh{\bb{C}}^{\dg} & =\wh{\bb{C}}^{\dg}\otimes1+\mathcal{U}^{-2}\otimes\bb{\wh{C}}^{\dg}+\frac{1}{2}\bb{H}\,\mathcal{U}^{-2}\otimes\bb{C}^{\dg}-\frac{1}{2}\bb{C}^{\dg}\otimes\bb{H}.\label{Y(g)}
\end{align}


\newcommand{\nlin}[2]{\href{http://xxx.lanl.gov/abs/nlin/#2}{\tt nlin.#1/#2}}
\newcommand{\hepth}[1]{\href{http://xxx.lanl.gov/abs/hep-th/#1}{\tt hep-th/#1}}
\newcommand{\arXivid}[1]{\href{http://arxiv.org/abs/#1}{\tt arXiv:#1}}
\newcommand{\Math}[2]{\href{http://xxx.lanl.gov/abs/math.#1/#2}{\tt math.#1/#2}}
\newcommand{\xmath}[1]{\href{http://xxx.lanl.gov/abs/math/#1}{\tt math/#1}}

\end{document}